\documentclass[12pt]{article}

\advance\voffset by -2.0cm \advance\hoffset by -1.25cm
\usepackage{pstricks,psfrag}
\usepackage{epsfig}
\usepackage{amsmath}
\usepackage{amssymb}
\usepackage{amsthm}
\usepackage{amsfonts}

\numberwithin{equation}{section}
\newtheorem{theorem}{Theorem}[]
\newtheorem{lemma}[theorem]{Lemma}

\theoremstyle{definition}

\newcommand{\CC}{\mathbb{C}} 
\newcommand{\RR}{\mathbb{R}} 
\newcommand{\QQ}{\mathbb{Q}} 
\newcommand{\ZZ}{\mathbb{Z}} 

\DeclareMathOperator{\im}{Im} 
\DeclareMathOperator{\Tr}{Tr} 
\DeclareMathOperator{\Sp}{Sp} 

\newcommand{\A}{\mathcal{A}} 
\newcommand{\D}{\mathcal{D}} 
\newcommand{\C}{\mathcal{C}} 

\newcommand{\N}{\mathcal{N}} 
\newcommand{\J}{\mathcal{J}} 
\newcommand{\Li}{\mathcal{L}} 
\newcommand{\Hh}{\mathcal{H}}
\newcommand{\M}{{\mathcal M}} 
\newcommand{\Sieg}{\mathfrak{H}} 
\newcommand{\Sch}{\mathfrak{S}} 
\newcommand{\be}{\begin{equation}}
\newcommand{\ee}{\end{equation}}
\newcommand{\OO}{\mathcal{O}}

\newcommand{\ie}{{\it i.e.~}}

\title{\vspace{-1cm}
\begin{flushright}{\small CALT-68-2774}\end{flushright}\vspace{1cm}
{\LARGE Genus two partition functions of chiral conformal field theories}
\vspace*{0.5cm}}

\author{
{\Large Matthias R.\ Gaberdiel}$^{1}$\thanks{\tt E-mail:
gaberdiel@itp.phys.ethz.ch}\ , {\Large Christoph A.\
Keller}$^2$\thanks{\tt E-mail: ckeller@theory.caltech.edu}
\vspace{0.2cm} \\ and \vspace{0.2cm} \\
{\Large Roberto Volpato}$^1$\thanks{\tt E-mail:
volpato@itp.phys.ethz.ch}
\\ \\
{$^1$ Institut f\"ur Theoretische Physik,
ETH Z\"urich} \vspace*{0.1cm} \\
{8093 Z\"urich, Switzerland} \vspace{0.8cm} \\
$^2$ {California Institute of Technology}\\
{Pasadena, CA 91125, USA} }
\date{\today}
\begin{document}
\maketitle

\begin{abstract}
A systematic analysis of the genus two vacuum amplitudes of chiral
self-dual conformal field theories is performed. It is explained that the
existence of a modular invariant genus two partition function implies
infinitely many relations among the structure constants of the theory.
All of these relations are shown to be a consequence of the associativity of the OPE,
as well as the modular covariance properties of the torus one-point functions.
Using these techniques we prove that for the
proposed extremal conformal field theories at $c=24k$ a
consistent genus two vacuum amplitude exists for all $k$, but that this does
not actually check  the consistency of these theories beyond what is already
testable at genus one.
\end{abstract}

\newpage
\renewcommand{\theequation}{\arabic{section}.\arabic{equation}}

\section{Introduction}\label{s:intro}

Usually, a 2d conformal field theory is defined by specifying the spectrum
of the theory (typically in terms of representations of some chiral algebra), as well
as the operator product expansions (OPEs) of the corresponding fields. The consistency
conditions require, in particular, that the OPE is associative, and that the theory
is modular covariant at genus $g=1$. Given these assumptions (or more precisely,
assuming that the polynomial relations of Moore \& Seiberg \cite{MS} are satisfied) the
theory is then also well-defined on higher genus Riemann surfaces.

There are, however, situations where a conformal field theory is characterised
in a different manner. In particular, in the context of the AdS$_3$/CFT$_2$ correspondence,
the gravity approach does not give access to the algebraic properties of the theory such as
its OPEs. Instead, we can only determine the vacuum amplitudes (at arbitrary genus)
from the gravity point of view \cite{Witten:2007kt,Maloney:2007ud,Gaiotto,Yin:2007gv,Yin:2007at}.
It is an old idea of Friedan \& Shenker \cite{Friedan:1986ua} that
a conformal field theory is also uniquely characterised in terms of these partition
functions (as functions of their modular parameters). Unlike the usual
approach to conformal field theory, this avenue has been much less explored. In
particular, there are two fundamental questions that are, to our knowledge,
still unanswered:
\begin{list}{(\arabic{enumi})}{\usecounter{enumi}}
\item Given the partition functions for all genera, is the conformal field theory
defined uniquely?
\item What are the consistency conditions a family of higher genus partition functions
    has to satisfy in order to define a consistent conformal field theory?
\end{list}
A little while ago, two of us \cite{GabVol} showed that knowing all genus partition functions
determines the current symmetry of the underlying conformal field theory
uniquely. Assuming a natural Lie algebraic conjecture, we could also show that
these amplitudes fix the representation content with respect to this current algebra
up to an overall automorphism of the Lie algebra. While this does not prove (1) --- for
example, these statements are vacuous for theories that do not have {\em any} currents ---
it gives very strong credence to it.
\smallskip

In this paper we shall begin to address the second question (2).
Given the complexity of higher genus amplitudes, we shall only be
able to explore the situation at genus $g=2$, and only for chiral
(self-dual) theories at $c=24k$. However, some of the salient
features are already visible there.  In particular, we shall show
that the existence of a modular invariant genus $g=2$ partition
function implies infinitely many relations among the structure
constants of the underlying conformal field theory;
this follows directly from the fact that the vector space of
such partition functions is finite dimensional.
As we shall
prove, all of those relations are a consequence of the associativity of the
OPEs (Jacobi identities of the ${\cal W}$-algebra), as well as
modular covariance at genus $g=1$, in nice agreement with the
analysis of Moore \& Seiberg. However, we can also identify at least
one additional consistency condition (beyond modular invariance)
that the vacuum amplitudes have to satisfy in order to define a
conformal field theory. This is the condition that the expansion
coefficients of the partition functions can actually be written in
terms of polynomials of individual structure constants. As we shall
show with two examples in section~5, this is a non-trivial
consistency condition which seems to go beyond modular invariance.
On the other hand, our analysis also suggests that this is {\em the
only} additional consistency condition.
\medskip

One of the main motivations for this work comes from the proposal of Witten \cite{Witten:2007kt}
regarding extremal conformal field theories. Witten proposed that the dual of pure gravity in
AdS$_3$ should be an extremal self-dual chiral conformal field theory with central charge
$c=24k$, $k\in {\mathbb Z}$, where $k$ is proportional to the AdS radius. Here `extremal'
means that the theory contains, apart from the Virasoro descendants of the vacuum,
only fields with conformal weight $h\geq k+1$. Self-duality implies, in particular, that the
character of the vacuum representation must be modular invariant by itself, and these
two conditions then fix the vacuum character (and hence the total partition function) completely.
For $k=1$ the extremal theory is the famous Monster theory, but the question of whether
the theories with $k\geq 2$ exist remains an open problem. Indeed, while the spectrum is
modular invariant (by construction), it is far from obvious whether one can define an associative
OPE on the corresponding set of fields. Using modular differential
equations, two of us  \cite{Gaberdiel:2007ve,Gaberdiel:2008pr} have argued that the theories
should be inconsistent for large $k$ ($k\geq 42$),  but unfortunately there is still a
small loophole in the argument.

For the extremal ansatz at $k=2$ and $k=3$, it was shown in
\cite{Witten:2007kt} (for $k=2$) and \cite{Gaiotto} that one can
define a consistent genus $g=2$ partition function. Given
that the existence of a modular invariant genus $g=2$ vacuum
amplitude implies infinitely many relations among the structure
constants of the theory (see above), this would appear to represent
a highly non-trivial consistency check for these theories. However,
as we shall also explain in this paper, this is somewhat misleading.
Indeed, many of the relations involve in fact coefficients that are
not otherwise known (and thus do not lead to any real
`constraints'), while the remaining `testable' relations (of which
there are still infinitely many!) turn out to be a consequence of the
associativity of the Virasoro algebra and the modular covariance
of certain simple 1-point functions at genus $g=1$. In fact, we can
prove rather generally (see Theorem 2
in section~4) that a `consistent' genus $g=2$ partition function
exists for a large class of putative theories, that include, in
particular, the extremal ansatz at arbitrary $k$. Furthermore, it
is clear from the assumptions of this Theorem that
the existence of a genus $g=2$ amplitude does not impose
any constraints beyond those that can already be analysed at genus one.
\bigskip

The paper is organised as follows. In section~2 we explain that the space of modular
invariant genus two vacuum amplitudes is finite dimensional for fixed  $c=24k$. We also
identify different sets of expansion coefficients that determine the genus two partition function
uniquely. In section~3 the various different expansion coefficients are interpreted from a
conformal field theory point of view. We identify the linear relations between
the different expansion coefficients that arise as a consequence of modular
invariance at genus two, and show that they are a consequence of the associativity of
the OPE and modular covariance at genus one. Some explicit examples are worked out in
section~3.3. In section~4 we apply these techniques to prove that a consistent genus two
vacuum amplitude exists for the extremal ansatz at arbitrary $k$. We also show that this
property is quite generic, and that it does not actually test any consistency conditions
beyond what is already testable at genus one --- see Theorem~2. Finally, in section~5 we
identify the additional consistency condition that has to be satisfied in order for the
vacuum amplitudes to define a consistent conformal field theory. We also
estimate the behaviour of this constraint at large genus, and suggest
that  it will eventually ({\it i.e.}\ for sufficiently large genus) become
very constraining for the extremal ansatz. Section~6 contains our conclusions, and
there are a number of appendices
where some of the more technical material has been collected.

\section{Genus two modular forms}\label{s:PFinvar}

As was for example explained in \cite{GabVol}, the genus two
partition function of a meromorphic conformal field theory at
central charge $c=24 k$ is of the form
\be\label{laZ}
Z_2 = \frac{W}{F^{12 k }} \ ,
\ee
where $W$ is a modular form of weight $12 k$, while $F^{12k}$ serves as a
reference partition function --- it describes the chiral contribution of $24 k$
uncompactified free bosons to the genus two partition function.
At genus $g=2$ the modular form $W$ may be taken to be a Siegel modular form
$W(\Omega)$, where $\Omega$ is the
Riemann period matrix of the corresponding Riemann surface. The
period matrices provide a parametrisation of the moduli space of
Riemann surfaces with respect to which modular transformations
assume a particularly simple form; for a more detailed explanation of all of this see
appendices~\ref{A:Schottky} and \ref{a:partfunct}.

\begin{figure}[h]
\begin{center}\resizebox{\textwidth}{!}{\input{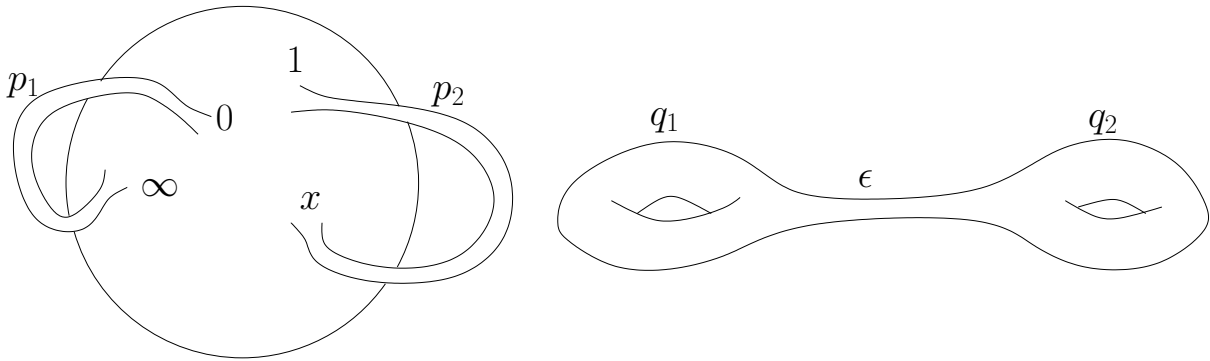}}
\end{center}
\caption{\label{fig:coord} To the left, the geometric interpretation
of the Schottky coordinates $p_1$ and $p_2$; the third coordinate
$x$ is given by the cross section of the insertion points. To the
right, the geometric interpretation of the sewn tori coordinates
$q_1$, $q_2$, $\epsilon$.}
\end{figure}

In order to analyse the factorisation properties of partition
functions under degeneration limits, however, other parametrisations
are more appropriate. The most basic one, which we will call the
`sewn-tori coordinates', comes from the so-called plumbing fixture
construction where one joins two tori (of modular parameters $q_1$
and $q_2$) by a cylinder whose modular parameter is described by a
third variable $\epsilon$, see figure~\ref{fig:coord}. This is the
parametrisation that was used in the work of Mason and Tuite
\cite{MT1,MT2,Tuite}.

Alternatively, we may use that any modular form $W(\Omega)$ may be
lifted to an automorphic form $\hat{W}$ on the Schottky space
$\Sch_2$ (that forms a finite covering of the moduli space of genus
two surfaces, see appendix \ref{a:Schottky}). The Schottky space
$\Sch_2$ can be identified with the open subset of $\CC^3$ defined
by
\be\label{schdomain}
\Sch_2:=\{(p_1,p_2,x)\in\CC^3\mid x\neq 0,1,\
0<|p_i|<\min\{|x|,\,1/|x|\}\ , i=1,2\}\ , \ee where the relation to
the usual Riemann period matrix $\Omega$ is
\be\label{coordrel}
e^{2\pi i \Omega_{11}} =p_1(1+{\cal O}(p_2))\ ,\qquad e^{2\pi i
\Omega_{22}} =p_2(1+{\cal O}(p_1))\ ,\qquad e^{2\pi i \Omega_{12}}=
x+{\cal O}(p_1 p_2)\ . \ee This parametrisation is appropriate to
describe the degeneration where the genus two surface becomes a
sphere with two thin handles connecting $0$ and $\infty$, and $x$
and $1$, respectively, see figure~\ref{fig:coord} and
appendix~\ref{A:Schottky}. The Schottky parametrisation has also
been applied to the analysis of higher loop string amplitudes, see
for example
\cite{DiVecchia:1988cy,DiVecchia:1989id,Frizzo:1999zx,Magnea:2004ai,Russo:2007tc,Duo:2007he}.

\medskip

\medskip

In the following we shall analyse the structure of  $Z_2$ as in
(\ref{laZ}), using just modularity and regularity properties. In
particular,  we shall take $W$ to be {\em any} Siegel modular form
of weight $12k$ (and $\hat{W}$ its lift to the Schottky space), and
we shall assume that $Z_2$ has smooth limits at the boundary of
moduli space. However, we shall not assume that the function $W$ (or
$Z_2$) arises from a consistent conformal field theory.

\subsection{Siegel modular forms of degree $g=2$}

At genus two, the space of Siegel modular forms of degree $g=2$ and
even weight is freely generated by
\be
\psi_4\ , \qquad \psi_6\ , \qquad \chi_{10}\ , \qquad \chi_{12}\ ,
\ee where the subscript denotes the modular weight (see appendix
\ref{a:modforms}). Since we are only interested in forms of weight
$12 k$, it is useful to introduce a set of generators for the
corresponding subring
\be
\psi_4^3\ ,\qquad \psi_{12}\ , \qquad \chi_{12}\ , \qquad
\chi^{(d)}_{12w_d} \ . \ee Here we have defined
$\psi_{12}=\frac{\psi_4^3-\psi_6^2}{1728}$, and $\chi^{(d)}_{12w_d}$
is the modular form of smallest weight $w=12 w_d$ with $w_d\in
{\mathbb N}$, that contains as a factor $\chi_{10}^d$. More
explicitly, we have $\chi^{(0)}_0=1$ as well as
\begin{align}
\chi_{24}^{(1)}&=\chi_{10}\psi_4^2\psi_6 &
\chi_{24}^{(2)}&=\chi_{10}^2\psi_4 \nonumber \\
\chi_{36}^{(3)}&=\chi_{10}^3\psi_6 &
\chi_{48}^{(4)}&=\chi_{10}^4\psi_4^2 \label{wddef} \\
\chi_{60}^{(5)}&=\chi_{10}^5\psi_4\psi_6 &
\chi_{60}^{(6)}&=\chi_{10}^6\ , \nonumber
\end{align}
and for arbitrary $d> 6$ we define recursively
\be\label{wddef1}
\chi^{(d)}_{12 w_{d}}=\chi_{10}^{6}\, \chi^{(d-6)}_{12(w_{d}-5)}\
,\qquad d>6\ . \ee Note that $w_d$ is given by the formula
\be\label{closewd}
w_d= d-\Big\lfloor\frac{d}{6}\Big\rfloor +\delta_{1,{d\bmod 6}}\ .
\ee The space of modular forms of $w=12 k$ is then spanned by the
modular forms
\be\label{e:gener}
\phi_{a,b,c,d}=\psi_4^{3a}\,\psi_{12}^b\,\chi_{12}^c\,\chi^{(d)}_{12w_d}
\ , \ee where $a,b,c,d$ are non-negative integers in the set
\be
{\cal P}_k = \{ (a,b,c,d) : a+b+c+w_d = k  \} \ . \ee The function
$Z_2$ in (\ref{laZ}) thus takes the form
\be\label{Z2gen}
Z_2 = \sum_{(a,b,c,d)\in {\cal P}_k} f_{abcd}
\frac{\hat\phi_{a,b,c,d}}{F^{12k}} = \sum_{(a,b,c,d)\in {\cal P}_k}
f_{abcd} \left(\frac{\hat\psi_4^3}{F^{12}}\right)^a \, \left(
\frac{\hat\psi_{12}}{F^{12}} \right)^b \,
\left(\frac{\hat\chi_{12}}{F^{12}}\right)^c \,
\frac{\hat\chi^{(d)}_{12w_d}}{F^{12w_d}} \ , \ee where $f_{abcd}$
are some constants, and $\hat\psi$ and $\hat\chi$ are the lifts of
$\psi$ and $\chi$, respectively, to the Schottky space. The various
factors have an expansion as
\begin{align}
\frac{\hat\psi_4^3}{F^{12}}&= 1+ 744(p_1+p_2)+\cdots\label{e:exppsi4}\\
\frac{\hat\psi_{12}}{F^{12}}&=p_1+p_2+\cdots\label{e:exppsi12}\\
\frac{\hat\chi_{12}}{F^{12}}&=p_1p_2\Bigl(1+\frac{1}{12}\frac{(x-1)^2}{x}\Bigr)+\cdots
\label{e:expchi12}\\
\frac{\hat\chi^{(d)}_{12w_d}}{F^{12w_d}}&=\frac{(x-1)^{2d}}{x^d}\Bigl(-\frac{1}{4}p_1p_2+\cdots\Bigr)^d
\ , \label{e:expchi10}
\end{align}
where the ellipses denote higher powers in $p_1$ or $p_2$. The
contribution of the term proportional to $f_{abcd}$ is thus
\be\label{e:expgen}
\frac{\hat\phi_{a,b,c,d}}{F^{12k}}=
\left(p_1^{b+c+d}p_2^{c+d}+p_1^{c+d}p_2^{b+c+d} \right) \Bigl(
\left(-{\textstyle \frac{1}{4}} \right)^d f(x)+\cdots\Bigr) \ , \ee
where $f(x)$ is explicitly given as
\be
f(x)=\frac{(x-1)^{2d}}{x^d} \sum_{l=0}^{c}  \binom{c}{l}
({\textstyle \frac{1}{12}})^l\,  \frac{(x-1)^{2l}}{x^{l}} \ . \ee

\subsection{Expansion coefficients}

The above argument shows that the space of $g=2$ modular forms of
weight $12 k$ is finite dimensional; indeed, the set ${\cal P}_k$ has
\be
|{\cal P}_k | = \frac{k^3}{5} + {\cal O}(k^2)
\ee
elements. In particular, this implies
that the genus two partition function $Z_2$ must be uniquely
determined in terms of a finite set of coefficients in some suitable
coordinate expansion. As we have mentioned above, there are two
different classes of coordinates that one may naturally use.

\subsubsection{Schottky expansion}

In the Schottky parametrisation the lift of $Z_2$ to the Schottky
space has the power series expansion
\be\label{ppexpans}
\hat{Z}_2=\frac{\hat W}{F^{12k}}=\sum_{h_2,h_1=0}^\infty
C_{h_1,h_2}(x)\, p_1^{h_1}p_2^{h_2}\ , \ee where $C_{h_1,h_2}(x)$ is
a rational function of $x$ whose only poles are at $0,1,\infty$; the
order of the poles are bounded by
\be\label{converge}
C_{h_1,h_2}(x)\stackrel{x\to 0}{\sim} {\cal O}(x^{-h_1-h_2})\ ,
\qquad C_{h_1,h_2}(x)\stackrel{x\to 1}{\sim} {\cal O}(1)\ , \qquad
C_{h_1,h_2}(x)\stackrel{x\to \infty}{\sim} {\cal O}(x^{h_1+h_2})\ .
\ee For example, the first of these properties can be proven by
restricting $\hat{Z}_2$ to the curve
$(p_1(t),p_2(t),x(t))\subset\Sch_2$, with
\be
x(t)=t\ ,\qquad p_1(t)=t/2\ ,\qquad p_2(t)=t/2\ , \ee where
$t\in\CC$ and $0<|t|<1/2$. The requirement that $\hat{Z}_2$ has a
finite limit as $|t|\to 0$ ({\it i.e.}\ at the boundary of moduli space)
then leads to the first bound in (\ref{converge}); the
other bounds can be proven similarly.

By construction $\hat{Z}_2$ is also modular invariant, and this
implies that the functions $C_{h_1,h_2}(x)$ must satisfy
\begin{subequations}\label{e:condiz}\begin{align}\label{e:condiza}
C_{h_1,h_2}(x)&= C_{h_2,h_1}(x) \ ,\\ \label{e:condizb}
C_{h_1,h_2}(x)&=C_{h_1,h_2}(1/x)\ .
\end{align}\end{subequations}
Indeed, the first identity comes from considering the modular
transformation that acts on the usual fundamental cycles
$\{\alpha_1,\alpha_2,\beta_1,\beta_2\}$ as
$(\alpha_1,\alpha_2,\beta_1,\beta_2) \mapsto
(\alpha_2,\alpha_1,\beta_2,\beta_1)$, while the second one comes
from the transformation $(\alpha_1,\alpha_2,\beta_1,\beta_2)\mapsto
(\alpha_1,\alpha_2^{-1},\beta_1,\beta_2^{-1})$. Because of the first
equation in \eqref{e:condiz} we may, from now on, consider only the
functions $C_{h_1,h_2}(x)$ with $h_2\le h_1$.
\smallskip

\begin{figure}[htb]\begin{center}
\resizebox{\textwidth}{!}{\input{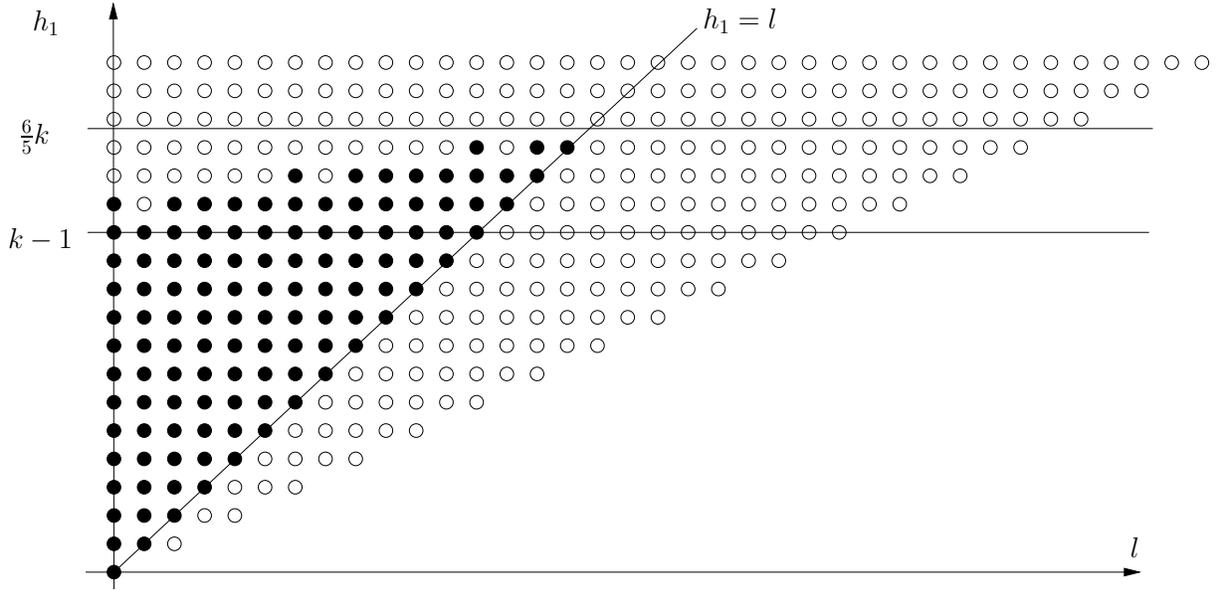}}\end{center}
\caption{\small A graphical representation of the set ${\cal
P}^{(*)}_k$ for $k=13$. Each (either white or black) circle in the
diagram denotes a pair $(h_1,l)$ for which we can find a $h_2\leq
h_1$ with $0\leq l \leq h_1+h_2$. Black circles denote pairs
$(h_1,l)$ for which at least one such choice of $h_2$ corresponds to
an element in ${\cal P}^{(*)}_k$. }\label{f:freepar}\end{figure}

For the following it will be useful to expand $C_{h_1,h_2}(x)$ in a
power series. A particularly simple expansion is
\be\label{gl3}
C_{h_1,h_2}(x)=\sum_{l=0}^{h_1+h_2} C^{(*)}_{h_1,h_2;l} \,
\frac{(x-1)^{2l}}{x^l} \ ,
\ee
since (\ref{converge}) and
(\ref{e:condizb}) imply that the sum on the right hand side is
finite. Since $W$ is a modular form of weight $12 k$, it is clear
that these coefficients can be expressed in terms of linear
combinations of the $f_{abcd}$ that appear in (\ref{Z2gen}), {\it
i.e.}\
\be
C^{(*)}_{h_1,h_2;l}  = \sum_{(a,b,c,d)\in {\cal P}_k}
M_{(h_1,h_2;l)}^{abcd} \, f_{abcd} \ ,
\ee
where $M$ is a matrix that
depends on $k$. We should furthermore expect that we can invert this
relation. However, since there are infinitely many coefficients of
the form $C^{(*)}_{h_1,h_2;l}$ --- for each fixed $h_1$ and $h_2$,
$l$ only takes the finitely many values $l=0,\ldots,h_1+h_2$, but
there are infinitely many values for $h_1$ and $h_2$ --- we need to
understand more precisely which of these coefficients are in fact
independent. By comparing (\ref{gl3}) with (\ref{e:expgen}) it is
easy to see that we may take the independent coefficients to be
labelled by
\be\label{P*k}
{\cal P}^{(*)}_k = \{ (h_1=b+c+d,h_2=c+d;l=d) :  b,c,d\in{\mathbb N}, \,
b+c+w_d \leq k  \} \ . \ee In particular, this then implies that we
can express the $f_{abcd}$ as
\be
f_{abcd} = \sum_{(h_1,h_2;l)\in{\cal P}^{(*)}_k}
\hat{M}_{abcd}^{(h_1,h_2;l)} \, C^{(*)}_{h_1,h_2;l} \ . \ee Note
that the set ${\cal P}_k^{(*)}$ consists of all triplets
$(h_1,h_2,l)$ of integers for which
\be\label{bound}
0\le l\le h_2\le h_1\le k+l-w_l \qquad \qquad h_1,h_2,l\in\ZZ\ . \ee
Obviously the inequalities \eqref{bound} only have a solution
provided that $w_l\leq k$. Because of (\ref{closewd}), we have the
bounds
\be\label{wbd}
\frac{5}{6} \,l \leq w_l \le l+1\ , \ee
 and thus a necessary condition for
$(h_1,h_2,l)\in {\cal P}_k^{(*)}$ is
\be\label{2.29}
(h_1,h_2,l)\in {\cal P}_k^{(*)} \quad \Longrightarrow  \quad 0\le
l\le h_2\le h_1\le k+\frac{l}{6} \leq \frac{6}{5} \, k\ , \ee where
in the last inequality we have used that $l\leq k+\frac{l}{6}$
implies $l\leq \frac{6}{5} k$. On the other hand, using the upper
bound from (\ref{wbd}) implies that a sufficient condition for
$(h_1,h_2,l)$ to be in ${\cal P}_k^{(*)}$ is
\be
0\le l\le h_2\le h_1\le k-1 \quad \Longrightarrow  \quad
(h_1,h_2,l)\in {\cal P}_k^{(*)} \ . \ee The structure of ${\cal
P}_k^{(*)}$ is sketched in figure~\ref{f:freepar}.

%

\subsubsection{Expansion in sewn tori coordinates}

Later on we shall also need the description of the genus two
partition function in terms of the sewn tori coordinates $q_1$,
$q_2$ and $\epsilon $, see figure~\ref{fig:coord}. With respect to
these coordinates we have an expansion as
\be\label{DRdef}
Z_2 = \sum_{h_1,h_2,l = 0}^{\infty} \D_{h_1,h_2;l}\, \, q_1^{h_1}\,
q_2^{h_2} \, \epsilon^{2l} \ . \ee The coefficients $\D_{h_1,h_2;l}$
are symmetric under the exchange of $h_1\leftrightarrow h_2$, and we
may therefore restrict ourselves to considering the terms with
$h_1\geq h_2$. The explicit relation between the sewn tori and the
Schottky coordinates is given by
\be
q_1 = p_1 \bigl(1 + {\cal O}(p_1,p_2) \bigr) \ , \quad
q_2 = p_2 \bigl(1 + {\cal O}(p_1,p_2) \bigr) \ , \quad
\epsilon  = (x-1)\bigl(1 + \OO(x-1)^2+ {\cal O}(p_1,p_2) \bigr) \ .
\ee
Obviously, again only finitely many of these coefficients are
independent, and we may take them to be  $\D_{h_1,h_2;l}$, with
$(h_1,h_2;l)$ in
\be\label{PDk}
{\cal P}^{(\D)}_k = \left\{ (h_1,h_2;l) : 0\leq l\leq h_2\leq
h_1\leq k+l-w_l  \right\} \ . \ee

\section{The conformal field theory perspective}\label{s:CFTpersp}

Up to now we have analysed the modular properties of the partition
functions $Z_2$, but we have not assumed that they arise from an
underlying conformal field theory. As we have explained in the
previous section there are infinitely many relations between the
expansion coefficients of $C_{h_1,h_2}(x)$ and the coefficients
$\D_{h_1,h_2;l}$ (since all of them are determined in terms of the
finitely many coefficients labelled by ${\cal P}_k$). These
relations encode constraints the underlying conformal field theory
has to satisfy in order to define a consistent genus two partition
function. In the following we want to exhibit these constraints more
explicitly.

\subsection{Invariants of the conformal field theory}

To start with we need to explain the conformal field theory
interpretation of the different expansion coefficients.

\subsubsection{The Schottky expansion}

In the Schottky parametrisation it is clear from the geometrical
definition (see appendix~C) that the coefficient functions
$C_{h_1,h_2}(x)$ have the interpretation
\be
C_{h_1,h_2}(x) = \sum_{\substack{\phi_1,\psi_1\in\Hh_{h_1}\\
\phi_2,\psi_2\in\Hh_{h_2}}}
G^{-1}_{\phi_1\psi_1}G^{-1}_{\phi_2\psi_2}\, \langle
V^{out}(\phi_1,\infty)V^{out}(\phi_2,x)V^{in}(\psi_2,1)V^{in}(\psi_1,0)
\rangle\ . \ee Here $G_{\phi \psi}$ is the metric on the space of
states (with $G^{-1}_{\phi\psi}$ the inverse metric), and the sums
over $\phi_j,\psi_j$ run over a basis of states at conformal weight
$h_j$. Finally, $V^{in}$ and $V^{out}$ are defined as in
\eqref{e:Vin} and \eqref{e:Vout}, and the 4-point correlator is
evaluated on the sphere. Note that the crossing symmetry of these
correlation functions implies directly (\ref{e:condiz}), see
eqs.~\eqref{symmh1h2} and \eqref{symm1x}. Furthermore, the
regularity conditions (\ref{converge}) are a consequence of the
property of the $L_0$-spectrum of the conformal field theory to be
bounded from below by zero.

It is convenient to restrict the sum over the states at conformal
dimension $h_2$ to the {\em quasiprimary} states $\Hh_{h_2}^{qp}$,
leading to the `quasiprimary functions'
\be\label{fourpprim}
\C_{h_1,h_2}(x)= (x-1)^{2h_2}\sum_{\phi,\psi\in\Hh_{h_2}^{qp}}
G^{-1}_{\phi\psi}\Tr_{\Hh_{h_1}}\bigl(V(\psi,1)V(\phi,x)\bigr)\ .
\ee
Because the contribution of the $L_{-1}$-descendant states can
be described in terms of differential operators, the general case is
then of the form
\be
C_{h_1,h_2}(x)=\sum_{h_2'\le h_2}K_{h_2,h_2'}(x)\, \C_{h_1,h_2'}(x)
\ , \ee where $K_{h_2,h_2'}$ are some differential operators (that
are independent of the theory in question). Any linear relations
between the coefficient functions $C_{h_1,h_2}(x)$ must therefore
come from linear relations between the quasiprimary functions
$\C_{h_1,h_2}(x)$. In the following we shall therefore restrict our
attention to these quasiprimary correlation functions. Since
$\C_{h_1,h_2}$ still satisfy \eqref{e:condizb} and \eqref{converge},
we can expand them as in \eqref{gl3}
\be\label{gl3p}
\C_{h_1,h_2}(x)=\sum_{l=0}^{h_1+h_2} \C^{(*)}_{h_1,h_2;l} \,
\frac{(x-1)^{2l}}{x^l} \ .
\ee
However, from the point of view of conformal field theory, the expansion,
\be\label{gl1p}
\C_{h_1,h_2}(x) = \sum_{l=0}^{\infty} \C_{h_1,h_2;l}^{(0)}
\,(x-1)^{2h_2} x^{l-h_1-h_2} \ ,
\ee
which converges for $|x|<1$, is
more natural. Indeed, while the conformal field theory
interpretation of the coefficients $\C_{h_1,h_2;l}^{(*)}$ is not
immediate, eq.\ \eqref{fourpprim} implies that
\be\label{C0def}
\C^{(0)}_{h_1,h_2;l}=\sum_{\phi,\psi\in\Hh_{h_2}^{qp}}
G^{-1}_{\phi\psi}\,
\Tr_{\Hh_{h_1}}\bigl(V_{-h_1+l}(\psi)V_{h_1-l}(\phi)\bigr)\ . \ee It
is clear that we can express the $\C^{(0)}_{h_1,h_2;l}$ coefficients
in terms of the $\C^{(*)}_{h_1,h_2;l}$ coefficients, and vice versa.
Indeed, we can rewrite (\ref{gl3p}) for any $L\geq 0$ as
\begin{eqnarray}
\C_{h_1,h_2}(x)& = & \sum_{l=0}^{L} \C^{(*)}_{h_1,h_2;h_1+h_2-l}
\, (x-1)^{2h_2}  \frac{(x-1)^{2h_1-2l}}{x^{h_1+h_2-l}} + {\cal O}(x^{L-h_1-h_2+1}) \nonumber \\
& = &  \sum_{l=0}^{L} \C^{(0)}_{h_1,h_2;l} \, (x-1)^{2h_2}
x^{l-h_1-h_2} +  {\cal O}(x^{L-h_1-h_2+1}) \ , \label{2.32}
\end{eqnarray}
from which we conclude that we can express the coefficients
\be\label{rel1}
\left\{ \C^{(0)}_{h_1,h_2;l}  : l =0,\ldots, L \right\} \quad
\hbox{as linear combinations of} \quad \left\{
\C^{(*)}_{h_1,h_2;h_1+h_2-l} : l =0,\ldots, L \right\}
\ee
and vice
versa. In particular, given the definition of ${\cal P}^{(*)}_k$,
this implies that the independent $\C^{(0)}_{h_1,h_2;l}$
coefficients can be taken to be labelled by
\be\label{P0k}
{\cal P}^{(0)}_k = \left\{ (h_1,h_2;h_1+h_2-l) : 0\leq l\le h_2\leq
h_1\leq k+l-w_l  \right\} \ . \ee The fact that only the invariants
with $l=0,\ldots,h_1+h_2$ can be independent is actually directly a
consequence of the crossing symmetry \eqref{e:condizb}.
Eq.~\eqref{rel1} holds for a general $L$ if we define
$\C^{(*)}_{h_1,h_2;l}=0$ for $l<0$ or $l>h_1+h_2$.

\subsubsection{The sewn tori coordinate expansion}
The conformal field theory interpretation of the coefficients
$\D_{h_1,h_2;l}$ in (\ref{DRdef}) is more immediate. In fact, it
follows directly from the geometric interpretation of the sewn tori
coordinates (see figure~\ref{fig:coord}) that the expansion
coefficients have the interpretation
\be\label{Dcft}
\D_{h_1,h_2;l}=\sum_{\phi,\psi\in\Hh_{[2l]}} G^{-1}_{\phi\psi}\,
\Tr_{\Hh_{h_1}}\bigl(V_0(\phi)\bigr) \,
\Tr_{\Hh_{h_2}}\bigl(V_0(\psi)\bigr)\ .
\ee
This was taken to be the starting point in the analysis of Mason \& Tuite
\cite{MT1,MT2}. Here $\Hh_{[h]}$ denotes the eigenspace (with
eigenvalue $h$) of the operator $L_{[0]}$, the scaling operator on
the torus, defined by
\be\label{e:L0def}
L_{[0]}=L_0+\sum_{n=1}^\infty\frac{(-1)^{n+1}}{n(n+1)}\, L_n\ .
\ee
Physically, this modified scaling operator comes from
the conformal transformation to torus coordinates.
Mathematically, its justification is that
the torus one point function $Z_\phi$ defined by
\be
Z_\phi(q)=\Tr_{\Hh}\bigl(V_0(\phi)q^{L_0-k}\bigr) \ , \qquad q= e^{2\pi i \tau} \ ,
\ee
with $\tau$ the modular parameter on the torus, is an
elliptic modular function of weight $h$, provided that $\phi$ has
$L_{[0]}$ eigenvalue $h$ \cite{Zhu}. Note that $L_0 = L_{[0]}$ on
Virasoro primary states.
\smallskip

This concludes our discussion of the conformal field interpretation of
the expansion coefficients.
For a consistent conformal field theory one also usually requires
that the theory has a unique $SL(2,\CC)$-invariant vacuum. This
condition puts some additional constraints on the above parameters.
In particular, it implies that
\be\label{Ndef}
\D_{h_1,h_2;0} = N_{h_1}\,N_{h_2} \ , \ee where $N_i=\dim \Hh_i$. As
we have seen in section~2, the independent coefficients
$\D_{h_1,h_2;l}$ are labelled by ${\cal P}_k^{(\D)}$ defined in
(\ref{PDk}). For $l=0$ the independent coefficients are thus
characterised by $0\leq h_2 \leq h_1 \leq k$. Because of
(\ref{Ndef}) this implies that the dimensions $N_i$ for $i\leq k$
are free parameters, while the dimensions $N_j$ with $j>k$ cannot be
independently chosen. This fits in perfectly with the constraints
obtained by imposing modular invariance of the genus $g=1$ torus
partition function \cite{Hoehn,Witten:2007kt} (see also the
discussion in the following section). In fact, the latter point of
view also implies that the invariants of the form $\D_{h_1,h_2;0}$
with $h_1>k$ depend only on the $\D_{h_1,h_2;0}$ with $h_2\le h_1\le
k$.

\subsection{Interpretation of the linear relations}

As we have seen in section~2, consistency of the conformal field
theory at genus $g=2$ implies that the invariants \eqref{C0def} and
\eqref{Dcft} must satisfy a large number of linear relations. On the
other hand, the analysis of Moore \& Seiberg \cite{MS} shows that
the consistency of a conformal field theory at higher genus is a
consequence of the associativity of the OPE on the sphere, as well
as the modular covariance of the torus one-point functions. Indeed,
for the self-dual theories we are discussing there, there is only
one irreducible representation, namely the vacuum representation
itself, and the fusing and braiding matrices are trivial. Thus the
polynomial relations of \cite{MS} are obviously satisfied provided
that the chiral algebra is itself consistent, {\it i.e.}\ that it
has an associative OPE. In addition, one has to require that the
one-point functions on the torus are modular covariant of
appropriate weight (so that the associated $S$ matrix is indeed
unity). Under these conditions the analysis of \cite{MS} implies
consistency at arbitrary genus. Thus the linear relations of the
invariants \eqref{C0def} and \eqref{Dcft} must be a consequence of
(1) associativity of the OPE and (2) modular covariance at genus
one.

In order to explain that this is indeed so we shall proceed as
follows. We shall first identify (see section~3.2.1) which relations follow
from the associativity of the OPE on the sphere, and which
are a consequence of the modular covariance of the torus one point
functions (see section~3.2.2). Then we shall show (see section~3.2.3)
that taking these relations
together allows us to relate any invariant of the theory
to a linear combination of
invariants associated to ${\cal P}_k^{(*)}$, thus mirroring the
relations that arise from modular invariance at genus $g=2$. To
prove that the two sets of relations are actually the {\em same}, we
note that the theta series of even self-dual lattices span the space
of modular forms of degree $g=2$ and weight $w=0$ mod $4$
\cite{Ozeki:1976,Bocherer:1983}. The corresponding lattice conformal
field theories are consistent conformal field theories at arbitrary
genus, and thus must satisfy both sets of constraints.
Since the constraints take the form of a system of linear
equations, it follows that the two sets
are in fact equivalent.

\subsubsection{Relations from the associativity of the OPE}

Let us begin with collecting the relations that follow from the
associativity of the OPE on the sphere. From the point of view of
conformal field theory the expansions \eqref{gl3p} and \eqref{gl1p}
are different expansions of a certain four point function on the
sphere, see eq.\ (\ref{fourpprim}). The consistency of the chiral
algebra implies, in particular, that these functions are
well-defined, irrespective of how one expands them. In turn, this
means that associativity of the OPE implies the relations
(\ref{rel1}). By a similar argument one can also show, assuming the
associativity of the OPE, that we can express the invariants
\eqref{Dcft} in terms of the invariants appearing in (\ref{gl3p}),
{\it i.e.}\ that
\be\label{rel2}
\bigl\{ \C^{(*)}_{h_1,h_2';l}:l=0,\ldots,L,\ h_2'\le h_2\bigr\}\
\text{ are linear comb. of }\ \bigl\{ \D_{h_1,h_2';l}:l=0,\ldots,L,\
h_2'\le h_2\bigr\} \ , \ee and vice versa; since this is slightly
technical, the proof is given in appendix~D.1. Associativity of the
OPE finally implies that the traces appearing in (\ref{C0def}) are
cyclic. In particular, we therefore have the identity
\begin{eqnarray}
\C^{(0)}_{h_1,h_2;l} & = & \sum_{\phi,\psi\in\Hh_{h_2}^{qp}}
G^{-1}_{\phi\psi}\, \Tr_{\Hh_{h_1}}\Bigl(V_{-h_1+l}(\phi)V_{h_1-l}(\psi)\Bigr) \nonumber \\
& = & \sum_{\phi,\psi\in\Hh_{h_2}^{qp}} G^{-1}_{\phi\psi}\,
\Tr_{\Hh_{l}}\Bigl(V_{h_1-l}(\psi)V_{-h_1+l}(\phi)\Bigr)
=\C^{(0)}_{l,h_2;h_1}\ .  \label{rel3}
\end{eqnarray}Altogether we therefore see that
associativity of the OPE implies the relations (\ref{rel1}),
(\ref{rel2}) and \eqref{rel3}.

\subsubsection{Relations from modular covariance of the torus one-point function}

Next we turn to the relations that follow from the modular
covariance of the torus one-point functions.
As was mentioned before in section~3.1.2, for each $\phi\in\Hh_{[h]}$,
{\it i.e.}\ each $\phi\in\Hh$ with $L_{[0]} \phi = h \phi$, the
one-point function on the torus $Z_\phi$ has modular weight $h$ and
is holomorphic for $\tau$ in the upper half plane, except for a pole
of order at most $k$ at $q=0$ \cite{Zhu}. We denote the space of
such functions by $M_{h,k}$. It is not difficult to show that
$M_{h,k}$ is isomorphic to the space $M_{12k+h,0}$ of (holomorphic)
modular forms of weight $12k+h$ (without any poles): indeed, given
an element in $M_{h,k}$, the corresponding modular form is obtained
by multiplication with $\Delta^k$, where $\Delta$ is the unique
modular
form of weight $12$ and leading term $\Delta(q) \sim q + {\cal O}(q^2)$.

The ring of modular forms is freely generated by the Eisenstein
series $G_4$ and $G_6$, and thus there are only modular forms of
even weight. The dimension of $M_{h,k}$, for even $h$, equals
\be \label{ddef}
d_{h,k}=\dim(M_{h,k})=\dim(M_{12k+h})=
k+\Big\lfloor\frac{h}{12}\Big\rfloor+1-\delta_{2,{h\bmod 12}}\ . \ee
For $h=2l$ this can be rewritten, using the definition of $w_l$ in
(\ref{closewd}), as
\be\label{3.12}
d_{2l,k}=k+l-w_l+1\ . \ee
\smallskip

\noindent It is convenient to choose a basis $\xi_l^{h,k}(q)$,
$l=0,\ldots,d_{h,k}-1$, of $M_{h,k}$ such that
\be
\xi_l^{h,k}(q)=q^{-k} \Bigl(q^l+\OO(q^{d_{h,k}}) \Bigr)\ . \ee Then
for any $\phi\in\Hh_{[h]}$, we can write the torus one-point
function as
\be\label{torusmod}
Z_{\phi}(q) = \sum_{n=0}^\infty
q^{n-k}\Tr_{\Hh_{n}}\bigl(V_0(\phi)\bigr)
=\sum_{l=0}^{d_{h,k}-1}\xi_l^{h,k}(q)\Tr_{\Hh_{l}}\bigl(V_0(\phi)\bigr)
\ .
\ee
In particular, this therefore implies that the first $d_{h,k}$
coefficients of the $q$-expansion of (\ref{torusmod}) determine all
remaining power series coefficients.

Given the definition of the invariants $\D_{h_1,h_2;l}$
(\ref{Dcft}), and by the obvious identity
$\D_{h_1,h_2;l}=\D_{h_2,h_1;l}$, it is therefore clear that the
invariants with $0\leq h_2\leq h_1 \leq d_{2l,k} -1$ determine all
other $\D_{h_1,h_2;l}$ invariants. Since $d_{2l,k}-1=k+l-w_l$
this means that we can always express
\be\label{rel4} \D_{h_1,h_2;l}\quad\text{ as a linear combination of
}\quad\Bigl\{\D_{h_1',h_2';l}: h_2'\leq h_1'\leq k+l-w_l\Bigr\}\ .\ee

\subsubsection{Completeness of these relations}\label{ss:completeness}

Finally, it remains to show that the conformal field theory
relations explained in sections~3.2.1 and 3.2.2 allow us to relate
any  invariant in terms of the invariants
$\C^{(*)}_{h_1,h_2;l}$ with $(h_1,h_2;l)\in {\cal P}^{(*)}_k$. The
precise statement is given by the following theorem.
\smallskip

\begin{theorem} \label{ThmComplete}
The relations \eqref{rel1}, \eqref{rel2}, \eqref{rel3} and
\eqref{rel4} are sufficient to express all invariants
$\C^{(0)}_{h_1,h_2;l}$ and $\D_{h_1,h_2;l}$, defined in
\eqref{C0def} and \eqref{Dcft}, as linear combinations of
$\C^{(*)}_{h_1,h_2;l}$ with $(h_1,h_2;l)\in {\cal P}^{(*)}_k$.
\end{theorem}
\smallskip

\noindent
The proof of this Theorem is given in appendix~\ref{app:Thm1}.
\medskip

Given the existence of the consistent lattice theories (see the
beginning of section~3.2) this demonstrates that associativity of
the OPE on the sphere and modular covariance of the torus
one-point functions is sufficient to guarantee the consistency of
the genus $g=2$ partition function, in agreement with the argument
of Moore \& Seiberg \cite{MS}.

\subsection{Examples and the contracted Jacobi identities}

In the previous section we have shown that the linear relations that
are required for the consistency of the genus $g=2$ amplitudes
follow from the associativity of the OPE on the sphere, as well as
from the modular covariance of the torus one-point functions.
However, it should be clear that the consistency at genus $g=2$ only
checks some partial aspects of these requirements. To see this
explicitly, it is useful to consider some simple examples.

\subsubsection{Self-dual theories at $c=24$}

For the case of $c=24$ we have $k=1$, and the set of free parameters
labelled by (\ref{bound}) are
\be
C^{(*)}_{0,0;0} \ , \qquad C^{(*)}_{1,0;0}\ , \qquad C^{(*)}_{1,1;0}
\ . \ee At $h=1$ all states are Virasoro (quasi-)primary, and thus
we need not distinguish between the quasiprimary functions
$\C_{h_1,h_2}(x)$ and the actual 4-point functions $C_{h_1,h_2}(x)$.

In terms of  the sewn tori coordinate expansion, the free
parameters are labelled by (\ref{PDk}), and are thus given by
\be
\D_{0,0;0}\ , \qquad \D_{1,0;0}\ , \qquad \D_{1,1;0}\ . \ee If we
assume that our conformal field theory has a unique vacuum, it
follows from (\ref{Ndef}) that we have
\be
\D_{0,0;0}=1\ , \qquad \D_{1,0;0}=N_1\ , \qquad \D_{1,1;0}=N_1^2\ ,
\ee where $N_1$ is the number of currents. This reproduces the
result of \cite{GabVol}, where it was shown that for $c=24$ the
genus $g=2$ amplitude is only a function of $N_1$, namely
\be
W = \psi_4^3+(N_1-744)\, \psi_{12}+(N_1-744)(N_1+984)\, \chi_{12}\ .
\ee
In particular, all the above invariants for a consistent
self-dual conformal field theory at $c=24$ must satisfy a polynomial relation of degree
two in the number of currents $N_1$.  To see how this works
explicitly, we note from \eqref{ppexpans} that the genus $g=2$
amplitude corresponding to $W$ has  the expansion
\be\begin{split}\hat Z_2=&1+N_1 (p_1+p_2)+196884
(p_1^2+p_2^2)\\&+p_1p_2\left[6 N_1+\frac{5 N_1^2}{6}+\left(-4
N_1+\frac{N_1^2}{12}\right)\Bigl( x+\frac{1}{x}\Bigr) +N_1
\Bigl(x^2+\frac{1}{x^2}\Bigr)\right] + \cdots \ .
\end{split}\ee
To obtain the invariants $C^{(0)}_{1,1;l}$ we expand the coefficient
of $p_1 p_2$ as
\be \label{ex11}
(x-1)^{-2}C_{1,1}(x)=\frac{N_1}{x^2}+\frac{-2
N_1+\frac{N_1^2}{12}}{x}+(N_1 + N_1^2) + 2 N_1^2 x + 3 N_1^2 x^2
+\cdots\ , \ee thus leading to
\begin{eqnarray}
\C^{(0)}_{1,1;0} & = & N_1\ , \qquad \qquad \;
\C^{(0)}_{1,1;1}  = -2 N_1+\frac{N_1^2}{12} \ , \\
\C^{(0)}_{1,1;2} &  = & N_1+N_1^2 \ ,  \qquad \C^{(0)}_{1,1;s} =
(s-1) N_1^2\ , \quad s\geq 3 \ . \nonumber \label{ex11c2}
\end{eqnarray}
On the other hand, it follows from the general structure of a
conformal field theory that the currents satisfy the commutation
relations
\be
{}[J^a_m,J^b_n] = i f_{abc} J^c_{m+n} + m \, \delta_{ab}\,
\delta_{m,-n} \ , \ee where $f_{abc}$ are the structure constants,
and we have normalised the currents (so that the central extension
term is proportional to $m$, rather than $km$.) In terms of these
structure constants one then finds
\begin{eqnarray}\label{ex11c1}
\C^{(0)}_{1,1;0} & = & \sum_a \Tr_{\Hh_1}(J^a_{-1}J^a_1) = N_1\ ,
\qquad \C^{(0)}_{1,1;1}  =  \sum_a\Tr_{\Hh_1}(J^a_0J^a_0)= -
\sum_{abc} f_{abc} f_{acb} \ ,
 \nonumber \\
\C^{(0)}_{1,1;2} &  = & \sum_a\Tr_{\Hh_1}(J^a_1J^a_{-1})=
\sum_a\Tr_{\Hh_1}(J^a_{-1}J^a_{1}) + \sum_a\Tr_{\Hh_1}([J^a_{1},J^a_{-1}]) = N_1+N_1^2 \ , \nonumber  \\
\C^{(0)}_{1,1;s} & = & \sum_a\Tr_{\Hh_1}(J^a_{s-1} J^a_{-s+1})=
 \sum_a\Tr_{\Hh_1}([J^a_{s-1}, J^a_{-s+1}]) = (s-1) N_1^2\ , \;\; (s\geq 3) \ .\qquad
\end{eqnarray}
Thus we obtain the non-trivial identity
\be\label{nontrivial}
\sum_{abc} f_{abc} f_{abc} = \frac{N_1^2}{12} - 2 N_1 \ , \ee where
we have used the total anti-symmetry of the structure constants
(which follows from the associativity of the OPE).

The analysis of the previous section implies that this identity must
follow from the associativity of the OPE, as well as the modular
covariance of the one-point functions. To see how this goes we
observe that modular covariance of the one-point functions implies
that
\be
Z_\phi = 0  \qquad \hbox{for} \qquad \phi = \sum_a J^a_{-1} J^a_{-1}
\Omega - \frac{N_1}{12} L_{-2} \Omega\ , \ee since $\phi$ is a
Virasoro primary state with $h=2$. Indeed, since $V_0(\phi)\Omega=0$
it follows that $Z_\phi$ must be holomorphic, but there are no
holomorphic modular forms of weight two. In particular, this
therefore implies that
\be
0 = \Tr_{\Hh_1}(V_0(\phi))=  - \frac{N_1^2}{12} +
\sum_a\Tr_{\Hh_1}(J^a_0J^a_0) + 2 \sum_a\Tr_{\Hh_1}(J^a_{-1}J^a_1) \
, \ee which is just (\ref{nontrivial}).

\subsubsection{Self-dual theories at $c=48$ and Jacobi identities}\label{sss:Jacobi}

For $c=48$, {\it i.e.}\ $k=2$, the free parameters can for example
be taken to be
\be
\D_{0,0;0} \ , \quad \D_{1,0;0}\ , \quad \D_{1,1;0} \ , \quad
\D_{2,0;0} \ , \quad \D_{2,1;0} \ , \quad \D_{2,2;0} \ , \quad
\D_{1,1;1} \ , \quad \D_{2,2;2} \ee or
\be
\C^{(0)}_{0,0;0} \ , \quad \C^{(0)}_{1,0;1}\ , \quad
\C^{(0)}_{1,1;2} \ , \quad \C^{(0)}_{2,0;2} \ , \quad
\C^{(0)}_{2,1;3} \ , \quad \C^{(0)}_{2,2;4} \ , \quad
\C^{(0)}_{1,1;1} \ , \quad \C^{(0)}_{2,2;2}  \ . \ee Let us for
simplicity assume that the theory does not have any currents,
$N_1=0$, and that the vacuum is unique. Then most of the above
coefficients are zero, and the only non-trivial parameters are
\begin{eqnarray}
& \D_{0,0;0} = 1 \ , \qquad & \D_{1,0;0}= 0 \ , \qquad \D_{1,1;0}=0
\,
\qquad \D_{2,0;0}=N_2 \ , \\
& \D_{2,1;0}=0 \ , \qquad & \D_{2,2;0} = N_2^2  \ , \qquad
\D_{1,1;1}=0 \ , \qquad \D_{2,2;2} ={d} \ ,
\end{eqnarray}
as follows from (\ref{Ndef}). Thus there are only two independent
parameters, namely $N_2$ and ${d}$. In terms of the $\C^{(0)}$
parameters, the only non-trivial invariants are
\begin{eqnarray}\label{3.37}
& \C^{(0)}_{0,0;0} = \Tr_{\Hh_0}(1)=1 \ , \qquad
& \C^{(0)}_{2,0;2}=\Tr_{\Hh_2}(1)=N_2\ , \\
& \C^{(0)}_{2,2;4}=\sum_j \Tr_{\Hh_2}(\hat{W}^j_{2} \hat{W}^j_{-2})
= b_1\ ,\qquad & \C^{(0)}_{2,2;2}=\sum_j \Tr_{\Hh_2}(\hat{W}^j_{0}
\hat{W}^j_{0}) = b_2 \equiv b\ , \nonumber
\end{eqnarray}
where $N_2=\dim(\Hh_2)$, and the $\hat{W}^j$ are an orthonormal
basis for the states at $h=2$. It follows that both $b_1$ and $b_2$
must be a function of $N_2$ and $d$; we can then take the
independent parameters also to be $N_2$ and $b$.
\smallskip

For a theory without currents we know that the commutation relations
of the $W^i_m$ modes take the form
(see for example \cite{Blumenhagen:1990jv})
\be \label{WWOPE}
[W^i_m,W^j_n] = \frac{c}{12} \delta^{ij} m (m^2-1) \delta_{m,-n} +
(m-n) h_{ijk} \, W^k_{m+n} + ig^{ij}_{\alpha} \, V^\alpha_{m+n}\ ,
\ee where $h_{ijk}$ and $g^{ij}_{\alpha}$ are structure constants,
and the $V^\alpha_n$ denote a basis for the space of states at
$h=3$. (The modes $W^j_m$ and $\hat{W}^j_m$ differ only by the
normalisation factor $\sqrt{2/c}$.) With these definitions we can
then calculate directly the 4-point functions\begin{eqnarray}\label{e:exp22}
\sum_{n,j} x^{-n-2}\Tr_{\Hh_2}\bigl(\hat{W}_{-n}^j
\hat{W}_{n}^j\bigr)& = &
\frac{2}{c} \, \sum_{n,j} x^{-n-2}\Tr_{\Hh_2}\bigl(W_{-n}^jW_{n}^j\bigr) \nonumber \\
& = & \frac{N_2}{x^4}+\frac{8H_1}{cx^2}+\frac{8H_2}{cx} + N_2 +
N_2^2 + \frac{16 H_2}{c} \cdots \ ,
\end{eqnarray}
as well as
\begin{align}\label{e:exp23}
\sum_{n,j}
x^{-n-2}&\Tr_{\Hh_3}\bigl(\hat{W}_{-n}^j\hat{W}_{n}^j\bigr)=
\frac{2}{c} \, \sum_{n,j} x^{-n-2}
\Tr_{\Hh_3}\bigl(W_{-n}^jW_{n}^j\bigr)=
\frac{4N_2}{x^5}+\frac{\frac{8}{c}H_1+\frac{4}{c^2}G}{x^3}+\cdots \ ,
\end{align}
where we have defined
\be\label{defhhf}
H_1=\sum_{ijk}h_{ijk}h_{ijk}\ , \qquad H_2=\sum_{ijk}h_{iij}h_{jkk}
\ , \qquad
G=\sum_{ij\alpha} g^{ij}_{\alpha} g^{ij}_{\alpha}\ .
\ee
As an aside we should note that some of these coefficients can obviously be calculated
in different ways which agree, however, if the Jacobi identity is satisfied. For example,
the coefficient proportional to $x^{-1}$ in (\ref{e:exp22}) was calculated as
\be
\sum_j\Tr_{\Hh_2}(\hat{W}^j_1 \hat{W}^j_{-1})=
\frac{2}{c} \, \sum_j\Tr_{\Hh_2}([W^j_1, W^j_{-1}])= \frac{8}{c} \sum_{ijk} h_{jji} h_{kki} \ .
\ee
However, we could have also evaluated it directly, by first  applying $W^j_{-1}$ to
the states $W^{k}_{-2}\Omega$ using
\be
W^j_{-1} \, W^k_{-2}\Omega = h_{jkm} W^m_{-3}\Omega
+ i g^{jk}_\alpha V^\alpha_{-3} \Omega \ ,
\ee
and then evaluating $W^j_1$ on the resulting state and taking the trace.
This would have led to a different expression for the coefficient of $x^{-1}$.
Requiring the two results to agree is precisely the identity (\ref{nontri}) below, which
is a consequence of the Jacobi identity.
\smallskip

\noindent
{}On the other hand, we know from (\ref{3.37}) that
\be\label{e:modexp22}
(x-1)^{-4}\C_{2,2}(x)=
\frac{N_2}{x^4}+\frac{b}{x^2}+\frac{N_2^2}{6x}+N_2+\frac{4
N_2^2}{3}+\frac{9 N_2^2}{2}x+\frac{32 N_2^2}{3} x^2+\cdots \ ,
\ee
as well as
\be \label{e:modexp23}
(x-1)^{-4}\C_{3,2}(x)= \frac{4
N_2}{x^5}+\frac{N_2^2}{6x^3}+\frac{15750720+240 b+369560
N_2}{x^2}+\frac{10746880 N_2+\frac{N_2^2}{6}}{x}+\cdots  \ . \ee By
comparing coefficients we can then obtain $H_1$, $H_2$ and $G$ as a
function of $N_2$ and $b$; in particular, we have from the
comparison of (\ref{e:exp22}) and (\ref{e:modexp22})
\be\label{relh12}
H_1 = \frac{cb}{8} \ , \qquad H_2 = \frac{c}{48} N_2^2\ ,
\ee
while the comparison of (\ref{e:exp23}) and (\ref{e:modexp23}) leads to
\be\label{relg}
G = \frac{c^2}{24} \, N_2^2 - \frac{c^2}{4}\, b \ .
\ee
For $c=48$ we then obtain the non-trivial relation
\be\label{nontri}
G=96\, (H_2-H_1)\ .
\ee
As pointed out above, this relation is actually a
consequence of the Jacobi identity of (\ref{WWOPE}). Indeed,
by considering the identity $[W^i_n,[W^j_m,W^k_l]]+ \textrm{cycl.} = 0$, we find the
constraint
\be \label{Jacobifh}
\frac{1}{2c} g^{jt}_\alpha g^{ki}_\alpha - h_{jks}h_{ist}+h_{ijs}h_{kst}=0\ .
\ee
Contracting with $\delta_{jk}
\delta_{it}$ we get
\be \label{JacobiInv}
\frac{1}{2c} (g^{ji}_\alpha)^2 - h_{jjs}h_{iis}+h_{ijs}h_{ijs} =
\frac{1}{2c} G + H_1 - H_2 =0\ .
\ee
Thus the consistency at genus
$g=2$ tests aspects of the Jacobi identity. It is clear, however,
that we only get identities between fully contracted expressions,
{\it i.e.}\ only between the specialised expressions
(\ref{JacobiInv}). Thus the genus $g=2$ amplitudes give us only {\em
partial} access to the full Jacobi identity (\ref{Jacobifh}).
This remark will be further explored in section~\ref{s:highgen}.

\section{Application to the extremal ansatz}\label{s:PFconstr}

Up to now we have studied the constraints a conformal field theory
has to satisfy in order to lead to a consistent genus two amplitude.
In particular, we have seen that consistency at genus two is
guaranteed if the OPE on the sphere is associative and the torus
one-point functions are modular covariant. However, as is apparent
from the arguments of section~3.2.2, these conditions are sufficient,
but certainly not necessary. Indeed, it is clear from the last
example above that only certain aspects of the associativity
constraints are actually relevant for the genus two consistency
analysis.

It is therefore an interesting question to ask the converse
question: how much information do we need about a conformal field
theory (be it ultimately consistent or not) in order to construct a
(seemingly) consistent genus two amplitude. This question is of
particular relevance in the context of the extremal theories
originally proposed in \cite{Witten:2007kt} as the dual conformal
field theory of pure gravity on AdS$_3$
\cite{Witten:2007kt,Maloney:2007ud}. This proposal was subsequently
modified for chiral massive gravity in  \cite{Li:2008dq}, but there
have been arguments to suggest that chiral gravity is actually
logarithmic
\cite{Grumiller:2008qz,Henneaux:2009pw,Maloney:2009ck,Skenderis:2009nt,Grumiller:2009mw},
in which case the dual conformal field theory would not be extremal
in the above sense. There is also a conformal field theory argument
that suggests that such extremal theories cannot exist for large
values of the central charge
\cite{Gaberdiel:2007ve,Gaberdiel:2008pr,Gaberdiel:2008xb}.
\smallskip

The proposed extremal conformal field theories are self-dual
meromorphic conformal field theories at $c=24 k$. Up to conformal
weight $h=k$ their spectrum only consists of Virasoro descendants of
the vacuum. In order for the torus partition function to be modular
invariant, the theory has to have additional primary states. These
appear only at conformal dimension $h>k$, and their number (and
conformal dimension) is uniquely fixed by the requirement that the
torus partition function is modular invariant.

In the context of the AdS$_3$/CFT$_2$ duality, the gravity calculation
gives in principle also access to the higher genus partition
functions, and one may ask whether their existence will be evidence
in favour of the consistency of these theories. In fact, for $k=2$
and $k=3$, the explicit genus $g=2$ partition function was
constructed in \cite{Witten:2007kt} and
\cite{Gaiotto,Yin:2007gv,Yin:2007at}. As we have seen in section~3,
the consistency of the genus two amplitudes checks certain aspects
of the associativity of the OPE, as well as the modular covariance of
the torus one-point functions. One may therefore expect that the
existence of these $g=2$ amplitudes provides a non-trivial
consistency check on the existence of these proposed theories.
Unfortunately, as we shall see in the following, this is not the
case.
\smallskip

In order to explain more precisely what we mean by this statement it
is instructive to consider a slightly more general situation.
Suppose that the spectrum of the putative self-dual chiral conformal
field theory is of the form
\be\label{ansatz}
\Hh = \Hh^{(0)} \oplus {\cal R} \ , \qquad {\cal R} \subset
\bigoplus_{n\geq B} \Hh_n \ , \ee where $\Hh^{(0)}$ is the vacuum
representation of some consistent chiral algebra ${\cal A}$, and ${\cal R}$ is
a (reducible) representation of ${\cal A}$. Here $\Hh_n$ is the
finite dimensional eigenspace of $\Hh$ with $L_0$-eigenvalue $n$,
and we assume that the $L_0$ spectrum of $\Hh$ is bounded from below
by zero, with a unique state, the vacuum, at conformal weight $h=0$.
Furthermore, we assume that the chiral algebra ${\cal A}$ contains
the Virasoro algebra at  $c=24 k$. On the other hand, the states in
${\cal R}$ all have conformal weight $h\geq B$, and we do not (a
priori) assume that we know anything about the OPE involving two
fields from ${\cal R}$. In the example of the extremal theories,
${\cal A}$ would be the Virasoro algebra, and the states in ${\cal R}$
describe the additional states whose conformal dimensions are
bigger or equal than $B=k+1$.

By assumption, the full space $\Hh$ is a representation of the
chiral algebra ${\cal A}$, and thus the torus one-point functions
\be\label{4.2}
Z_\phi = \Tr_{\Hh}\bigl(V_0(\phi)q^{L_0-k}\bigr)
\ee
with $\phi\in \Hh^{(0)}$, are determined by our ansatz (\ref{ansatz}). If $\phi$
has $L_{[0]}$ eigenvalue $h$, then we know on general grounds that
$Z_\phi$ must be an element of $M_{h,k}$, see the discussion in
section~3.1.2. Thus consistency at genus one already requires that
these torus one-point functions are modular invariant. Provided that
this is the case (as we shall from now on assume) the relations that
come from the modular covariance of the torus one-point functions
(see section~3.2.2) are therefore automatically satisfied.

In order to understand what constraints the genus two analysis
implies we therefore consider the other invariants we introduced
above, namely $\C^{(0)}_{h_1,h_2;l}$ and $\D_{h_1,h_2;l}$. Since
we do not know anything about the OPEs involving general states in
$\Hh$, we shall only consider the invariants that
can be defined and computed given the ansatz \eqref{ansatz}.
These include, in particular,
\be \label{inv1}
\C^{(0)}_{h_1,h_2;l} =
\sum_{\phi,\psi\in\Hh_{h_2}^{qp}}G^{-1}_{\phi\psi} \,
\Tr_{\Hh_{h_1}}\bigl(V_{l-h_1}(\phi)V_{h_1-l}(\psi)\bigr)\ , \qquad
h_2<B\ , \ee since for $h_2 < B$ the relevant states $\phi$ and
$\psi$ are part of the chiral algebra, and we know, by assumption,
how they act on the full space $\Hh$. Similarly, regarding the
invariants $\D_{h_1,h_2;l}$, we can calculate those for which
$h_2<B$ or $2l<B$. Indeed, in either case we can restrict the sum
over $\Hh_{[2l]}$ in
\be\label{inv2}
\D_{h_1,h_2,l}=\sum_{\phi,\psi\in\Hh_{[2l]}} G^{-1}_{\phi\psi}\,
\Tr_{\Hh_{h_1}}\bigl(V_0(\phi)\bigr) \,
\Tr_{\Hh_{h_2}}\bigl(V_0(\psi)\bigr)\ , \qquad h_2<B\quad \text{or} \quad 2l<B
\ ,\ee to states in $\Hh^{(0)}$. This is obvious for the case that
$2l<B$. For $h_2<B$ note that if  $\psi$ is not an $\A$ descendant
of the vacuum, $\Tr_{\Hh_{h_2}}V_0(\psi)$ vanishes since $\Hh_{h_2}$
contains only vacuum descendants. Thus in either case only $\A$
descendants contribute, whose contributions we can compute. Thus, we
may take \eqref{inv2}, with the sum restricted to $\Hh_{[2l]}\cap
\Hh^{(0)}$, as the definition of $\D_{h_1,h_2;l}$ for $h_2<B$ or
$2l<B$.

We conclude that there are infinitely many invariants, namely
\eqref{inv1} and \eqref{inv2} for arbitrary $h_1$, that  we can
calculate from our ansatz. On the other hand, we know from the
analysis of section~2 that consistency at genus two implies that
there are only finitely many free parameters, and thus that these
invariants must satisfy infinitely many linear relations. It would
thus seem that we get strong constraints from the consistency of the
genus two amplitudes!
\smallskip

Unfortunately, this impression is somewhat deceptive. As we have
seen in section~3 the constraints that arise in this way test
effectively aspects of the associativity of the OPE. However, since
we only have access to the invariants of the form (\ref{inv1}) with
$h_2< B$, we will only be able to test the associativity of the OPE
of the fields up to conformal weight $B$. But since we assumed that,
up to conformal weight $B$, the conformal field theory consists just
of elements in the chiral algebra ${\cal A}$ (for which
associativity is assumed) all of these constraints will be
automatically satisfied. More precisely we can prove the following:

\begin{theorem}\label{mainThm}
Suppose $\Hh$ is of the form \eqref{ansatz} and all torus one-point
functions $Z_\phi$ with $\phi\in \Hh^{(0)}$ are modular covariant.
Then there exists a (not necessarily unique) modular form of degree
$g=2$ and weight $12k$, whose coefficients in the expansions
\eqref{gl3} agree with \eqref{inv1} for $h_2 < B$, and whose
coefficients in the expansion \eqref{DRdef} with either $h_2<B$ or
$2l<B$ agree with
\eqref{inv2}.
\end{theorem}

\noindent {\bf Proof:}
The proof is analogous to Theorem 1, but this time we are not
allowed to assume that $\Hh$ is a vertex operator algebra. This
means that we cannot assume that {\em all} identities \eqref{rel1}, \eqref{rel2},
\eqref{rel3} and \eqref{rel4} hold. However, we can still
follow the same strategy as before: using those
identities that still follow from the associativity of ${\cal A}$ and the modular covariance
of the one-point functions (\ref{4.2}),
we can demonstrate that all invariants \eqref{inv1} and \eqref{inv2}
can be expressed in terms of linear combinations of some $\C^{(*)}_{h_1,h_2;l}$ with
$(h_1,h_2;l)\in{\cal P}^{(*)}_k$. Given our previous analysis we know, on the other hand,
that these relations are the same as those that arise from the modular covariance
at $g=2$, and hence the result follows.
\smallskip

To see this in more detail we first
note that the limits (\ref{converge}) still hold
since the $L_0$ spectrum is bounded by assumption.
Next, using that all operators in $\A$ are local, we write
\begin{multline}
(x^{-1}-1)^{2h_2}\Tr_{\Hh_{h_1}}(V(\psi,1)V(\phi,x^{-1}))
=x^{-2h_2}(x-1)^{2h_2}\Tr_{\Hh_{h_1}}(V(\phi,x^{-1})V(\psi,1))\\
=x^{-2h_2}(x-1)^{2h_2}\sum_n x^{n+h_2}\Tr_{\Hh_{h_1}}( \phi_n  \psi_{-n})
=(x-1)^{2h_2}\Tr_{\Hh_{h_1}}(V(\phi,1)V(\psi,x))\ ,
\end{multline}
from which \eqref{e:condizb} follows after summing over all states
in $\Hh_{h_2}$.
We
can thus again define invariants $\C^{(*)}_{h_1,h_2;l}$, $h_2<B$,
which are related to the $\C^{(0)}_{h_1,h_2;l}$ as in \eqref{rel1}.
Eq.~\eqref{rel3} is still valid for $h_2<B$ because it only makes
use of the cyclicity of the trace. Furthermore, as shown in appendix
\ref{app:CD}.1, \eqref{rel2} is valid for $h_2 < B$. Using the same
arguments as in the proof of Lemma~1 (see appendix \ref{app:Thm1})
it then follows that the invariants \eqref{inv1} and the invariants
\eqref{inv2} for $h_2<B$ can be correctly reproduced by a suitable
choice of the free parameters in
\be
S = \{ (h_1,h_2;l)\in {\cal P}^{(*)}_k : h_2 < B \}\ .
\ee

It remains to prove that the modular form can be consistently chosen
in such a way that also the invariants $\D_{h_1,h_2;l}$ with
$2l<B\le h_2$ are correctly reproduced. This works because the only
consistency checks of those invariants come essentially from the
modular covariance of the one point functions. More precisely, since
by assumption the one point functions are modular covariant, we can
use (\ref{rel4}) to express all such invariants by
$\D_{h_1',h_2';l}$ with $h_2'\le h_1'\le k+l-w_l$. If all $h_2'<B$,
we know from the arguments given above that the invariants are
consistent with all other invariants in (\ref{inv1}) and
(\ref{inv2}). If on the other hand, some $h_2'$ appears with
$h_2' \geq B$, then $2l<B$ implies $(h_1',h_2';l)\in{\cal P}_k^{(*)}$,
and thus we can simply fix the corresponding
$\C^{(*)}_{h_1',h_2';l}$. Thus the only effect of the invariants
(\ref{inv2}) with $2l<B\le h_2$ is to fix the free parameters
\be
S'=\{ (h_1,h_2;l)\in {\cal P}^{(*)}_k : 2l<B\le h_2 \}
\ .
\ee
This completes the proof.\qed

\bigskip

Note that in general $S\cup S'$ is a proper subset of ${\cal P}^{(*)}_k$, in which
case, the genus $g=2$ amplitude is not uniquely fixed by the
information on $\Hh$, {\it i.e.}\ by (\ref{inv1}) and
(\ref{inv2}). (An example of this is described below.) Finally, we observe that the
assumptions of the theorem can be weakened,
because modular covariance of $Z_\phi$ is only needed for all
$\phi\in \Hh^{(0)}$ with conformal weight less than $2B$.

\subsection{The extremal ansatz}

The above proposition can be directly applied to the proposed
extremal theories. As was already mentioned before, in this context
${\cal A}$ is the Virasoro algebra, and $B=k+1$. The proposition
then implies that a consistent genus two amplitude can be
constructed provided that the torus one-point functions $Z_\phi$ are
modular invariant (with the appropriate modular weight).

For the case at hand, this latter condition is actually
straightforward to check. Because the chiral algebra ${\cal A}$ is
just  the Virasoro algebra, the only states $\phi$ that appear are
Virasoro descendants of the vacuum. Using the recursion relations of
Zhu \cite{Zhu} (see also \cite{Gaberdiel:2008pr}) it is clear that
each such one-point function can be written as a modular covariant
differential operator $D_q^{(\phi)}$  acting on the vacuum torus
amplitude,
\be
Z_\phi(q)\equiv \Tr_{\Hh}\bigl(V_0(\phi)q^{L_0-k}\bigr)=D_q^{(\phi)}
\Tr_{\Hh}\bigl(q^{L_0-k}\bigr)\ , \qquad\
\phi=L_{[-2]}^{n_2}L_{[-3]}^{n_3}\cdots\Omega\ . \ee For example,
for $\phi=L_{[-2]}\Omega$, we have
\be
Z_{L_{[-2]}\Omega}(q)
\equiv (2\pi i)^2\Tr_{\Hh}\bigl((L_{0}-k)q^{L_0-k}\bigr)= (2\pi i)^2
q\frac{d}{dq}Tr_{\Hh}\bigl(q^{L_0-k}\bigr)\ , \ee so that
\be
D_q^{(L_{[-2]}\Omega)}=(2\pi i)^2 q\frac{d}{dq}\ . \ee If $\phi$ has
$L_{[0]}$ eigenvalue $h$ with $h$ even --- for $h$ odd, the torus
one-point function, and thus the differential operator, vanishes
identically --- the differential operator is of order $\tfrac{h}{2}$
and of modular weight $h$. Thus the modular covariance of the
one-point functions $Z_\phi$ is a direct consequence of the modular
invariance of the torus vacuum amplitude (which is satisfied by
construction).

This argument therefore demonstrates that a consistent genus $g=2$
amplitude, satisfying {\em all} relations (\ref{inv1}) with $h_2\leq
k$, exists for all $k$. This conclusion is obviously in agreement
with the results for $k=2$ and $k=3$ in
\cite{Witten:2007kt,Gaiotto}. However, it also shows that one is not
actually testing any non-trivial consistency conditions of the
extremal ansatz in this way. In fact, the Theorem shows that the
same would have been true for {\em any} ansatz satisfying the above
conditions.

\subsubsection{Explicit examples: the extremal ansatz for low values of $k$}

It is maybe instructive to see how the general machinery works out
for these examples. Let us first consider the case with $k=2$
($c=48$). The extremal ansatz is a special case of the example
studied in section~3.2.2. Indeed, if at $h=2$ we only have the
Virasoro field, the parameter $N_2$ and $B$ take the form
\be
N_2 = 1 \ , \qquad B = \frac{2}{c} \Tr_{\Hh_2}(L_0^2) =
\frac{8}{c}=\frac{1}{6}\ . \ee Thus it follows from (\ref{relh12})
that $H_1 = 1= H_2 = 1$, which is indeed in agreement with the
definition in (\ref{defhhf}).
\smallskip

For $k=3$ ($c=72$) the analysis is similar. In this case the free
parameters may be taken to be $N_1$, $N_2$, $N_3$, as well as
\be
\C^{(0)}_{1,1;1}\ , \quad \C^{(0)}_{2,1;2}\ , \quad
\C^{(0)}_{2,2;3}\ , \quad \C^{(0)}_{2,2;2}\ ,\quad \C^{(0)}_{3,2;3}\
, \quad \C^{(0)}_{3,3;4}\ , \quad \C^{(0)}_{3,3;3}\ . \ee All of
these invariants can be computed from the extremal ansatz, so that
the genus $2$ partition function is uniquely determined. We have
also checked explicitly that the low-lying invariants in
(\ref{inv1}) and (\ref{inv2}) are indeed correctly reproduced (as
our general argument predicts). We have also done the same analysis
at $k=4$.

Something new happens at $k=5$ since there is a modular form
$\chi_{10}^6$ of weight $12k=60$ whose leading term is of order
$p_1^6p_2^6$ in the expansion \eqref{e:expgen}. Its coefficient is
associated with the invariant
\be \C^{(0)}_{6,6;6}=
\sum_{\phi,\psi\in\Hh_6^{qp}} G_{\phi\psi}^{-1}
\Tr_{\Hh_6}\bigl(V_0(\phi)V_0(\psi)\bigr)
\ee
that cannot be determined directly from the extremal ansatz. Thus for
$k\geq 5$ the extremal ansatz does not specify the genus $g=2$ partition
function uniquely,\footnote{This corrects a statement in \cite{Gaiotto}.} and
there is a whole vector space of genus $g=2$ partition functions that
reproduce correctly all computable invariants.

\section{Remarks about higher genus}\label{s:highgen}

Let us close this paper by coming back to the question that
was raised in the introduction. Recall that according to an old
idea of Friedan and Shenker \cite{Friedan:1986ua}, the
higher genus partition functions determine a conformal
field theory uniquely. Assuming that this idea is correct, then constructing
all higher genus vacuum amplitudes would be a way of {\em defining}, say,
the dual conformal field theory of some gravity theory on AdS$_3$.
The interesting question would then be: what consistency conditions
apart from modular invariance do the higher genus vacuum amplitudes
have to satisfy in order to define a consistent conformal field theory?
\smallskip

Obviously, it is in practice quite hard to calculate these higher genus
amplitudes explicitly (see however \cite{Yin:2007at}), but one could
attempt to construct the higher genus amplitudes by symmetry
principles. For example, one could specify the algebra of low lying
states of the conformal field theory, and simply attempt to find
modular invariant
higher genus amplitudes that satisfy all appropriate factorisation
conditions with respect to these low lying states. At genus $g=2$,
Theorem~2 shows that this will be possible provided that one chooses
${\cal A}$ and ${\cal R}$ appropriately (where the required
conditions are rather weak). While we have not done any detailed analysis
at higher genus (see however section 5.2), it seems plausible that
a similar statement to Theorem~2 could also hold at higher genus.

Suppose then that we have found such a family of higher genus
amplitudes, say for the extremal ansatz. Would this prove that the
extremal theories exist?
As we have seen in section~3, modular invariance at genus
$g=2$ implies a large number of relations between different expansion coefficients
which, in terms of the conformal field theory, translate into the
statement that certain quadratic expressions in the structure
constants have to be equal. These in turn are a consequence of the
Jacobi identity of the ${\cal W}$-algebra, {\it i.e.}\ of the associativity
of the OPE, see section~4.1.  One would similarly expect that higher
genus constraints will give rise to relations which, from the point
of view of the underlying conformal field theory, involve higher
powers of these structure constants (which again should be implied
by the associativity of the OPE). Consistency at arbitrary genus would
thus probably imply that all possible contractions of the Jacobi identities
will be satisfied. Would this be sufficient to reconstruct the Jacobi
identities themselves, {\it i.e.}\ to prove associativity of the OPEs?

The answer to this question is no --- for a rather simple reason.
To explain this, let us briefly return to the example of section~\ref{sss:Jacobi}.
The higher
genus vacuum amplitudes give us access to fully contracted polynomials
of the structure constants, such as $H_1$, $H_2$ and $G$ in (\ref{defhhf}).
However, it is not guaranteed that we can find  individual structure
constants $h_{ijk}$ and $g^{ij}_{\alpha}$ (that must be
$N_2\times N_2\times N_2$ tensors and $N_2\times N_2\times N_3$ tensors,
respectively), whose contractions reproduce the given values of $H_1$, $H_2$ and
$G$. (Here $N_h$ denotes the number of states at conformal weight $h$,
which can be read off from the torus amplitude.) We shall demonstrate in the
following that this is a {\em non-trivial} consistency condition which does not seem
to be implied by the modular invariance of the higher genus amplitudes.
Thus a family of modular invariant higher genus vacuum amplitudes can only
define a consistent conformal field theory if this consistency condition is satisfied.
It is natural to speculate that this will be the only additional consistency condition
beyond modular invariance.

\subsection{Obstructions at $c=24$}\label{s:c24n123}

The simplest example where the above consistency condition is non-trivial appears
for the self-dual theories at $c=24$. As was shown in \cite{GabVol},
at $c=24$ consistent $g=1,2,3,4$ vacuum amplitudes (that have the
correct modular and factorisation properties) can be found for any
number of currents $N_1$. However, it is believed that
only $71$ consistent conformal field theories exist at $c=24$.
In particular, no theories exist for $0 < N_1 <24$ \cite{Schellekens}.
Unless $N_1$ is one of the special values for
which a consistent theory exists, there must be an obstruction
towards reconstructing the theory from the vacuum amplitudes.
Obviously, it is conceivable that this obstruction will manifest
itself in that one cannot find vacuum amplitudes (with the correct
modular and factorisation properties) for $g\geq 5$, but this seems
unlikely to us. Instead we believe that the obstruction appears in
that one cannot reproduce the coefficients of the modular forms in terms
of contracted structure constants. In fact, we can see this obstruction very
explicitly in some simple examples, as we shall now show.

In the following the states at $h=1$ will play an important role.
One knows on general grounds (see {\it e.g.}\ \cite{meromorphic})
that the modes of these fields satisfy the commutation relations
\be \label{algebra}
{}[J^a_m,J^b_n]=m\kappa^{ab}\delta_{m,-n}+if^{ab}_cJ^c_{m+n}\ , \ee
where $\kappa^{ab}$ is non-singular with inverse $\kappa_{ab}$.
Furthermore, $f^{ab}_c$ is antisymmetric in $a\leftrightarrow b$.

As is explained in \cite{GabVol} the number of currents $N_1$ fixes
the genus $g=2,3,4$ amplitudes uniquely. In turn, one can read off
from this that (at $c=24$)
\be\label{e:conditio}
\Tr_{\Hh_1}\Bigl(( \kappa_{ab}
J^a_0J^b_0)^l\Bigr)=N_1\left(\frac{N_1}{24}-1\right)^l\ , \quad
\hbox{for} \quad l=1,2,3\ . \ee We shall use this relation in the
following to show that the theories with $N_1=1,2$ (that do not
arise among the $71$ theories of \cite{Schellekens}) are in fact
inconsistent.

\subsubsection{The theory with $N_1=1$}

The simplest  case is the theory with $N_1=1$. It is immediately
clear that this theory is inconsistent since at $N_1=1$ the Lie
algebra (\ref{algebra}) is necessarily abelian (since $f^{ab}_{c}$
is antisymmetric in $a\leftrightarrow b$ and must hence vanish). As
a consequence the trace in (\ref{e:conditio}) vanishes for
$l=1,2,3$, which disagrees with the right hand side.

\subsubsection{The theory with $N_1=2$}

The situation with $N_1=2$ is more interesting. Since $\kappa_{ab}$
is symmetric, we may choose a basis of $\Hh_1$ that diagonalises
$\kappa_{ab}$, and set
\be
\kappa_{11}= \epsilon_1 \ ,\qquad \kappa_{22}=\epsilon_2\ , \qquad
if^{12}_1=\alpha\ ,\qquad if^{12}_2=\beta\ . \ee Then, $J^1_0$ and
$J^2_0$ in the adjoint representation correspond to the matrices
\be
J^1_0=\begin{pmatrix} 0 & \alpha\\ 0 & \beta
\end{pmatrix} \quad \hbox{and} \quad
J^2_0=\begin{pmatrix} -\alpha & 0\\ -\beta & 0
\end{pmatrix}
\ee and we find
\be
\kappa_{ab}J^a_0J^b_0=\begin{pmatrix} \epsilon_2\alpha^2 &
\epsilon_1\alpha\beta\\ \epsilon_2\alpha\beta & \epsilon_1\beta^2
\end{pmatrix}\ .
\ee The eigenvalues of this matrix are $0$ and
$\epsilon_2\alpha^2+\epsilon_1\beta^2$, so that \eqref{e:conditio}
becomes
\be
\Tr_{\Hh_1}\Bigl(( \kappa_{ab} J^a_0J^b_0)^l\Bigr)
=(\epsilon_2\alpha^2+\epsilon_1\beta^2)^l =2(-11/12)^l\quad
\hbox{for} \quad l=1,2,3  \ . \ee It is easy to see that this does
not have any solution.
\medskip

One can similarly analyse the situation with $N_1=3$, but there the
constraints from $l=1,2,3$ ({\it i.e.}\ from genus $g\leq 4$) are
not sufficient to lead to a contradiction. This is not surprising:
at $N_1=3$ there are four free parameters, namely the three
eigenvalues of $\kappa$, as well as the one totally anti-symmetric
structure constant $f^{123} = f^{12}_{c} \kappa^{3c}$. On the other
hand, \eqref{e:conditio} only gives rise to three equations, and  a
solution can be found. Indeed, one can take the currents to define
an affine $su(2)$ algebra at level $k$ with $k=-\frac{16}{7}$ since
then
\be
\kappa_{ab} J^a_0J^b_0 = - \frac{7}{8} \, {\bf 1}_{3\times 3} \quad
\hbox{on $\Hh_{1}$.} \ee This then solves \eqref{e:conditio}.
However, one would expect that this ansatz will not be compatible with
the vacuum amplitudes at $g\geq 5$.

\subsection{Polynomial constraints from modular invariance}

The analysis of the previous section shows that not every family of
modular invariant genus $g$ amplitudes defines a consistent conformal
field theory. Indeed, the main additional condition seems to be that there
is a solution for the individual structure constants such that the contracted
powers reproduce the expansion coefficients of the corresponding
modular amplitudes. Obviously, this condition only becomes powerful
at sufficiently high genus when the number of equations for the contracted
structure constants exceeds the number of (unknown) structure constants.
\smallskip

To be more specific, let $Z_g$ be a modular form of genus $g$ and weight $12k$, and
consider a  series expansion of $Z_g$ in $3g-3$ suitable parameters
$t_1,\ldots,t_{3g-3}$
\be\label{gengexp}
Z_g=\sum_{h_1,\ldots,h_{3g-3}}
C_{h_1,\ldots,h_{3g-3}}t_1^{h_1}\cdots t_{3g-3}^{h_{3g-3}}\ .
\ee
More precisely, let us consider a sphere with $2g$ punctures and
decompose it into a set of pair of pants, \ie into $2g-2$ spheres
with three punctures each. We then connect the punctures to obtain
the $g$ handles of the surface, and take the $t_i$ to parametrise the
$3g-3$ different tubes. For example, the possible decompositions of a genus $2$
surface are shown in figure \ref{fig:decompos}. As we shall see below, the number
of pant decompositions grows very quickly with $g$.
\begin{figure}[h]
\begin{center}\resizebox{\textwidth}{!}{\input{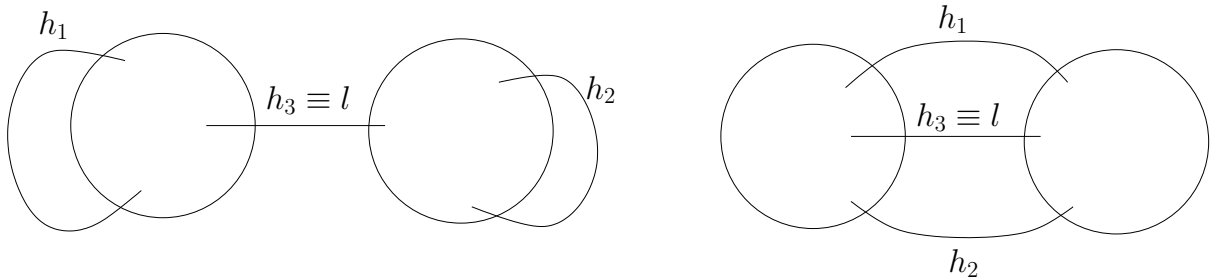}}
\end{center}
\caption{\label{fig:decompos} The two possible pants decompositions
of a genus $2$ surface. They correspond to the expansions in the
coordinates $q_1,q_2,\epsilon$ (left) and $p_1,p_2,x$ (right), and
the associated invariants are $\D_{h_1,h_2;l}$ (left) and
$\C^{(0)}_{h_1,h_2;l}$ (right).}
\end{figure}

If $Z_g$ is the genus $g$ partition function of a conformal field
theory, then for every pant decomposition,
each of the coefficients $C_{h_1,\ldots,h_{3g-3}}$ equals a
contracted combination of structure constants of the underlying
conformal field theory. (Different pant decompositions of a Riemann
surface correspond to different ways of contracting the indices.)
Alternatively, instead of working with
structure constants, we can also consider directly the $3$-point
functions of the conformal field theory. In this language the
coefficients $C_{h_1,\ldots,h_{3g-3}}$ must then equal contracted
products of  $3$-point functions where the relevant fields have
conformal weights $h_1,\ldots,h_{3g-3}$. For the following it is
convenient to define, for each fixed $L$, the set
\be
I_L(g):=
\{(h_1,\ldots,h_{3g-3}):0\le h_1,\ldots,h_{3g-3}<L\}\ .
\ee
Then the coefficients $C_{h_1,\ldots,h_{3g-3}}$ associated to
$I_L(g)$ are polynomials in the $3$-point functions of fields of weight
less than $L$ only. We will call the number of such $3$-point functions
$K_L$; it depends on the dimensions of the $L_0$-eigenspaces (that can
be read off from the torus partition function $Z_1$), but not on
$g$ nor on any other details of the theory. It is clear that there is a non-trivial
consistency condition of the above type if
\be\label{conscheck}
D(g)\left| I_L(g)) \right| >K_L\ , \ee where $D(g)$ is the number of
inequivalent pant decompositions of a genus $g$ surface. Obviously,
the left hand side grows to infinity as $g\rightarrow \infty$, while
the right hand side is independent of $g$. Thus, for sufficiently
large $g$, there will be many such consistency conditions. However,
if the relevant spaces of states are large (as is, for example, the
case for the extremal theories) one has to go to rather large values of
$g$ before one obtains non-trivial constraints in this manner.
\bigskip

If we are given a family of higher genus partition functions, then this is all
we can say. However, in the context of the extremal ansatz, the natural
question is slightly different. Suppose, as for Theorem~2, that our
putative conformal field theory is of the form (\ref{ansatz}). Can we
then construct a family of higher genus amplitudes that not only reproduce
the computable coefficients (as in Theorem~2), but also allow for a solution
of all expansion coefficients in terms of individual structure constants?

{}From this point of view, the unknown parameters are the free parameters
${\cal P}_k(g)$ labelling a genus $g$ modular form of weight $12 k$, as well
as the $3$-point functions involving  three fields from ${\cal R}$.
The problem is now that the set of coefficients ${\cal P}_k(g)$ also
grows with $g$ and, {\it a priori}, these free parameters could
allow one to adjust the vacuum amplitudes so that one can always solve
for the individual $3$-point functions (involving fields from ${\cal R}$).
However, this is not the case. As is shown in
appendix  \ref{a:slope}, for every $g$, ${\cal P}_k(g)$
does not contain any element $(h_1,\ldots,h_{3g-3})$ with
\be\label{pkbound} h_1,\ldots,h_{3g-3}> L(g)\equiv \frac{k}{5}(g+4)\ .
\ee
[This is the generalisation to arbitrary genus of the inequality (\ref{2.29}).]
Let us define, for given $M>0$, the set
\be
I_{L(g),M}(g):=\{(h_1,\ldots,h_{3g-3}):L(g)<
h_1,\ldots,h_{3g-3}<L(g)+M\}\ . \ee We want to consider the
subsystem of equations that require that the coefficients labelled
by $I_{L(g),M}(g)$ in all pant decompositions can be expressed in
terms of the $3$-point functions. Because of \eqref{pkbound},  these
equations are unaffected by our ambiguity in the definition of the
vacuum amplitudes, {\it i.e.}\ they are independent of the
parameters in ${\cal P}_k(g)$. Furthermore, for large $g$, $L(g)>B$,
and all the relevant $3$-point functions involve mostly fields from
${\cal R}$.

With these preparations we can now estimate the number of equations,
and the number of unknowns. To leading order in $g$, the number of
equations for each pant decomposition goes as
\be
\left| I_{L(g),M}(g) \right| \sim M^{3g-3}\ . \ee
A rough estimate
of the number of different pant decomposition is%
\footnote{See \cite{Bender:1978,Bollobas:1983,Chae:2005} for related
asymptotic formulae in graph enumeration problems.}
\be \label{Npairpants}
D(g)\sim \frac{(6g-6)!}{2^{3g-3}(3g-3)!\,(3!)^{2g-2}(2g-2)!}\
\stackrel{g\to \infty}{\sim}\
(2\pi(g-1))^{-1/2}\Bigl(\frac{3g-3}{2e}\Bigr)^{g-1} \ , \ee where
the right hand side is a lower bound on the number of double cosets
in the space
$(S_2^{3g-3}\times S_{3g-3})\backslash S_{6g-6}/(S_3^{2g-2}\times S_{2g-2})$,
with $S_n$ the group of permutations of $n$ elements. In
fact, each coset corresponds to a decomposition of either a
connected or a disconnected surface, but it is easy to see that the
contribution of the disconnected surfaces is of order $D(g)/g$ and hence
can be ignored in our leading order approximation. Thus, the number of
equations in the subsystem relative to $I_{L(g),M}(g)$ is
\be \#\text{ equations}\sim g^{-1/2}\Bigl(\frac{M^3(3g-3)}{2e}\Bigr)^{g-1}\
.\ee The number of unknowns is roughly speaking the number of
structure constants up to weight $L(g)+M$, so that\footnote{
Actually, we are overestimating the number of unknowns, because we
are also including the structure constants of $\A$, but this does
not affect the following reasoning.}
\be\label{numunkn} \#\text{ unknowns}\sim
\Bigl(\sum_{h\le L(g)+M}\dim\Hh_{h}\Bigr)^3=\Bigl(\sum_{h\le L(g)+M}n_{h}\Bigr)^3\ .\ee
Here, $n_{h}$ are the coefficients in the $q$-expansion of the
modular invariant torus partition function
\be
\frac{Z_1(q)}{\Delta^{k}}=\sum_{h=0}^\infty n_hq^{h-k}\
,\qquad q=e^{2\pi i\tau}\ ,
\ee
where $\Delta=q+\OO(q^2)$ has modular
weight $12$. For large $h$ the Cardy formula
\be
\label{numstates} n_{l+k}\sim \frac{k^{1/4}}{\sqrt
2}\;l^{-3/4}\;e^{4\pi\sqrt{kl}}\ ,\qquad\qquad l\gg 0\ \ee applies
(see appendix \ref{a:cardy} for details). Note that, at leading
order in $h\to \infty$, $n_h$ only depends on the modular weight
$12k$ of $Z_1$ and on the normalisation $n_0=1$. Thus, taking just
the leading contribution $h=L(g)+M$ in the sum \eqref{numunkn}, the
requirement that the system of equations is overdetermined is
\be\label{firstcons}
g^{-1/2}\Bigl(\frac{M^3(3g-3)}{2e}\Bigr)^{g-1}
\gtrsim\bigl(L(g)+M-k\bigr)^{-9/4}e^{12\pi\sqrt{k(L(g)+M-k)}}\ , \ee
up to some factor independent of $g$ and $M$. Now, if we take $M>0$
fixed and $g$ large enough so that $L(g)\sim kg/5\gg M-k$,
eq.~\eqref{firstcons} becomes
\be\label{roughes}
g^{-1/2}\Bigl(\frac{M^3(3g-3)}{2e}\Bigr)^{g-1}\gtrsim
g^{-9/4}\bigl(e^{\frac{12\pi k}{\sqrt 5}}\bigr)^{\sqrt{g}}\ , \ee
which is satisfied for sufficiently large $g$. Thus there will be
many obstructions to the construction of a family of modular
invariant genus $g$ partition functions if we demand that their
coefficients can be expressed in terms of contracted structure
constants.

To see how this estimate works in practice, let us consider the
example of the extremal ansatz at $k=2$. The estimate
\eqref{roughes} indicates that, for $M<5$, the first constraints on
the structure constants should arise for $20\lesssim g\lesssim 40$.
A more precise computation, using \eqref{Npairpants} and
\eqref{numunkn} with the correct values for $n_h$, shows that
consistency of the partition function at genus $g=23$ puts
non-trivial constraints on the structure constants up to conformal
weight $h=L(g)+M=17$. In practice, however, amplitudes of such a high
genus are unfortunately not accessible.

\section{Conclusions}

In this paper we have analysed the structure of the genus $g=2$
vacuum amplitudes of chiral self-dual conformal field theories. In particular,
we have shown that the existence of a modular invariant $g=2$ partition function
implies infinitely many relations among the structure constants of the theory.
All of  these relations are a consequence of the associativity of the OPE, as
well as the modular covariance of the torus one-point amplitudes (Theorem~1).
This was to be expected from the point of view of the Moore \& Seiberg
analysis \cite{MS}.

We have applied these techniques to the extremal ansatz, and we have
shown that a (seemingly) consistent genus $g=2$ vacuum amplitude
exists for the extremal theories at $c=24 k$ for all $k$ (Theorem~2). However,
we have  also shown that this does not check any consistency conditions
of the extremal ansatz beyond what can already be tested by analysing the
modular covariance properties of the ansatz at genus one.

Finally, we have shown that a family of modular invariant genus $g$ partition functions
can only define a consistent conformal field theory if one can actually
reconstruct the individual structure constants whose contracted expressions reproduce
the expansion coefficients of the vacuum amplitudes. As we have demonstrated
with two examples (see section~5.1) this is a non-trivial consistency condition that
does not seem to be implied by modular invariance. The rough counting argument
of section~5.2 suggests that it will also lead to a stringent constraint for the
extremal ansatz. However, as is also clear from that analysis, this constraint
will only become interesting at rather large genera --- and hence is unfortunately,
technically fairly out of reach. Thus to prove or disprove the existence of
the extremal ansatz by studying higher genus partition functions seems to be
as difficult as the brute force approach of analysing the Jacobi identities of the
fields at large conformal weight.

\section*{Acknowledgements}

We thank Terry Gannon for a useful conversation and subsequent correspondence.
We are particularly grateful to Ida Zadeh for pointing out
sign mistakes in appendix C of a previous version of this article.
C.A.K.\ thanks the Pauli Center for support during his visit to ETH.
The research of M.R.G.\ is partially supported by the Swiss
National Science Foundation, while the research of R.V.\ is supported by
an INFN Fellowship.

\appendix

\renewcommand{\theequation}{\Alph{section}.\arabic{equation}}

\section{Riemann surfaces, Schottky groups and modular forms}\label{A:Schottky}

\subsection{Riemann period matrices and modular
forms}\label{a:Riemann}

In order to analyse the modular properties of partition functions,
it is useful to define the period matrix of a Riemann surface. Let
$\Sigma$ be a compact Riemann surface of genus $g>0$. Let us define
a basis of the first homology group $H_1(\Sigma,\ZZ)$
$\{\alpha_1,\ldots,\alpha_g,\beta_1,\ldots,\beta_g\}$, with
symplectic intersection matrix
\be\label{symplbasis}
\#(\alpha_i,\alpha_j)=0=\#(\beta_i,\beta_j)\ ,\qquad
\#(\alpha_i,\beta_j)=\delta_{ij}\ ,\qquad i,j=1,\ldots,g\ . \ee This
condition determines the basis up to a symplectic transformation
\begin{equation}\label{modulhomol}
\Biggl(\begin{matrix}\alpha\\
\beta\end{matrix}\Biggr)\mapsto 
\Biggl(\begin{matrix}\tilde\alpha\\
\tilde\beta\end{matrix}\Biggr):=\Biggl(\begin{matrix}D & C \\ B &
A\end{matrix}\Biggr) \Biggl(\begin{matrix}\alpha\\
\beta\end{matrix}\Biggr)\ ,\qquad\qquad \Biggl(\begin{matrix}A & B
\\ C & D\end{matrix}\Biggr)\in \Sp(2g,\ZZ) \ , \end{equation}
where $\alpha$ and $\beta$ are $g$-dimensional vectors, and
$A,B,C,D$ are $g\times g$ matrices. The choice of such a basis
uniquely determines a basis $\{\omega_1,\ldots,\omega_g\}$ of
holomorphic $1$-differentials normalised with respect to the
$\alpha$-cycles
\be\label{e:modnorm}\oint_{\alpha_i}\omega_j=\delta_{ij}\ ,\qquad
i,j=1,\ldots,g\ . \ee The Riemann period matrix of $\Sigma$ is then
defined by
\be
\Omega_{ij}=\oint_{\beta_i}\omega_j\ , \ee and it has the properties
\be
\Omega_{ij}=\Omega_{ji}\ ,\qquad \im\Omega>0\ . \ee Obviously, the
basis $\{\omega_1,\ldots,\omega_g\}$, and the Riemann period matrix
depend on the choice of the symplectic basis of $H_1(\Sigma,\ZZ)$;
under the action \eqref{modulhomol} of the symplectic group, the
holomorphic 1-differentials and the period matrix transform as
\begin{subequations}
\begin{align}\label{e:modomega}(\omega_1,\ldots,\omega_g)&
\mapsto(\tilde\omega_1,\ldots,\tilde\omega_g)
=(\omega_1,\ldots,\omega_g)(C\Omega+D)^{-1}\ ,\\
\Omega&\mapsto \tilde\Omega=(A\Omega+B)(C\Omega+D)^{-1}\ .
\end{align}
\end{subequations}
Let us define the Siegel upper half-space as the space of $g\times
g$ symmetric complex matrices with positive definite imaginary part,
\be
\Sieg_g=\{Z\in M_g(\CC)\mid Z_{ij}=Z_{ji} , \, \im Z >0\}\ . \ee The
locus $\J_g\subseteq \Sieg_g$ of all the period matrices of genus
$g$ Riemann surfaces is dense in $\Sieg_g$ for $g\le 3$, whereas for
$g>3$ it is a $(3g-3)$-dimensional subspace of $\Sieg_g$. The
quotient $\J_g/Sp(2g,\ZZ)$ is isomorphic to $\M_g$; in particular,
the Riemann period matrices of two different Riemann surfaces lie in
different $Sp(2g,\ZZ)$-orbits in $\J_g$.

\subsection{Modular forms of degree $1$ and $2$}\label{a:modforms}

A (Siegel) modular form $f$ of degree $g$ and weight $k$ is a
holomorphic function on $\Sieg_g$ such that
\be
f\bigl((AZ+B)(CZ+D)^{-1}\bigr)=\det(CZ+D)^kf(Z)\ ,\quad
M=\Biggl(\begin{matrix}A & B
\\ C & D\end{matrix}\Biggr)\in \Sp(2g,\ZZ)\ .
\ee
For $g=1$ we also require that $f$ is holomorphic at the cusps;
a cusp is a fix-point $p\in\RR\cup\{\infty\}$ under the action of
some $M\in\Sp(2,\ZZ)\cong SL(2,\ZZ)$ with $\Tr(M)=\pm 2$ (a
parabolic element). An analogous condition is automatically
satisfied for $g>1$.

\bigskip

The space of modular forms of degree $1$ is generated by the
Eisenstein series $G_4$ and $G_6$ defined by
\be
G_{k}(q)=1+\frac{2}{\zeta(1-k)}\sum_{n=1}^\infty
\sigma_{k-1}(n)q^n\ ,
\ee
where $\zeta$ is the Riemann $\zeta$-function, and
\be \sigma_k(n)=\sum_{d|n}d^k\ .
\ee
All Siegel modular forms of degree $2$ can be written in terms
of Eisenstein series \be E_k:=\N_k\sum_{C,D}\det(C\Omega+D)^{-k}\ ,
\ee
where $\N_k$ is a normalisation constant, and $C$ and $D$ are $2\times 2$
integral matrices such that $\left(\begin{smallmatrix} A & B\\ C & D
\end{smallmatrix}\right)\in Sp(4,\ZZ)$. The sum is over all the
inequivalent pairs under left multiplication by elements of
$GL(2,\ZZ)$ \cite{VDG,Igusa}. The Eisenstein series admits a Fourier
expansion (\cite{VDG}, pages 17-18)
\be
E_k=\sum_{n,m=0}^\infty\sum_{\substack{r\in\ZZ\\
r^2\le4nm}}a_k(n,m,r)\,q_{11}^n\,q_{22}^m\,q_{12}^r\ , \ee where
\be\label{multper} q_{ij}:=e^{2\pi i\Omega_{ij}}
\ , \ee and
\be a_k(n,m,r)=\frac{2}{\zeta(3-2k)\zeta(1-k)}\sum_{d|(n,m,r)}d^{k-1}H\Bigl(k-1,\frac{4nm-r^2}{d^2}\Bigl)
\ . \ee Here, $H$ is Cohen's function (see \cite{EichZag}, pages
21-22 for a definition) 
 and the normalisation is chosen so that
$a_k(0,0,0)=1$.

The ring of (even) Siegel modular forms of degree two is freely
generated by
\begin{align}
\psi_4& =  E_4\ ,\qquad &\chi_{10} &= \frac{43867}{2^{12}\cdot
3^{5}\cdot 5^2\cdot 7\cdot 53}(E_{10}-E_4E_6) \ , \\
\psi_6&=E_6 \ ,\qquad
&\chi_{12} &= \frac{131\cdot 593}{2^{13}\cdot 3^{7}\cdot 5^3\cdot
7^2\cdot 337}(21^2E_4^3+250E_6^2-691E_{12}) \ .\notag
\end{align}

\subsection{Schottky parameters}\label{a:Schottky}

The Schottky uniformisation describes a general non-singular Riemann
surface as the quotient of the Riemann sphere
$\hat\CC=\CC\cup\{\infty\}$ by a suitable subgroup of $PSL(2,\CC)$.
Geometrically, a surface $\Sigma$ of genus $g>0$ is obtained by
cutting $2g$ disks from $\hat\CC$, bounded by non-intersecting
circles $C_1,\ldots,C_g,C_{-1},\ldots,C_{-g}$, and then by
identifying each circle $C_i$ with the circle $C_{-i}$ via a
suitable fractional linear transformation $\gamma_i\in PSL(2,\CC)$
such that \be\label{schcircle} \gamma_iC_{-i}=C_i\ ,\ee for all
$i=1,\ldots,g$. The discrete subgroup $\Gamma$ with distinct free
generators $\gamma_1,\ldots,\gamma_g$ is called a marked Schottky
group. Each $\gamma\in\Gamma$ is characterised by two distinct
points $a_\gamma,r_\gamma\in\hat\CC$ (called the \emph{attractive}
and \emph{repelling} fixed point, respectively) and a complex number
$p_\gamma\in\CC$ (the \emph{multiplier}), with $0<|p_\gamma|<1$,
such that
\be
\frac{\gamma(z)-a_\gamma}{\gamma(z)-r_\gamma}= p_\gamma\,
\frac{z-a_\gamma}{z-r_\gamma}\ ,\qquad\qquad \text{for all
}z\in\hat\CC\ .
\ee
 The Riemann surface $\Sigma$ can be obtained as
the quotient of $\hat\CC$ by $\Gamma$, and every non-singular closed
surface can be obtained in this way.
\begin{figure}
\begin{center}\resizebox{.6\textwidth}{!}{\input{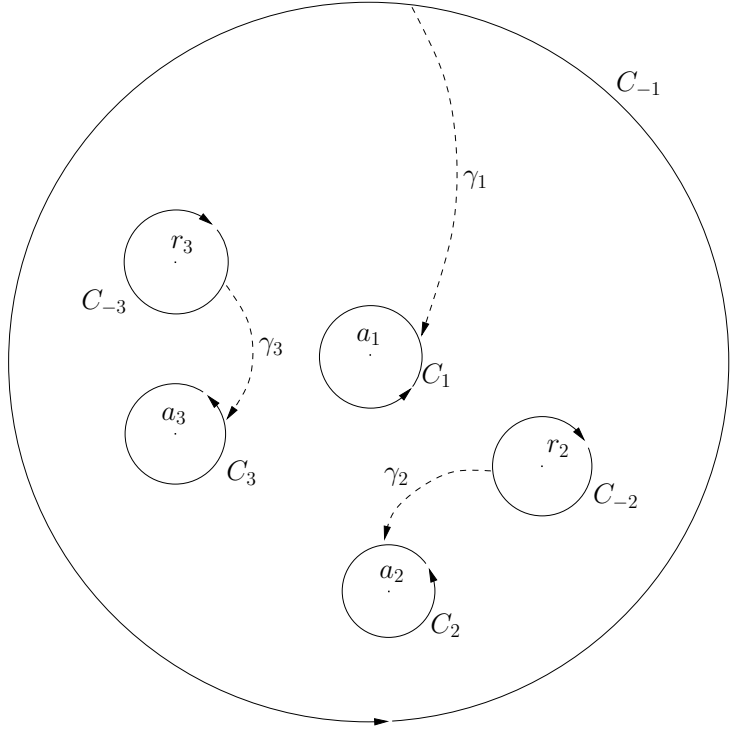}}
\end{center}
\caption{\label{fig:schottky} A fundamental domain for a Schottky
group. Each generator $\gamma_i$, $i=1,\ldots,g$, with fixed points
$a_i,r_i$, maps the circle $C_{-i}$ to the circle $C_{i}$.}
\end{figure}
By a $PSL(2,\CC)$ conjugation, one can fix
\be\label{normSch} a_1=0\ ,\qquad r_1=\infty\ ,\qquad a_2=1\ ,
\ee where we set \be a_{i}:= a_{\gamma_i}\ ,\qquad r_{i}:=
r_{\gamma_i}\ ,\qquad p_i:= p_{\gamma_i}\ .
\ee
When \eqref{normSch}
holds, the marked Schottky group is called \emph{normalised}. The
space $\Sch_g$ of normalised marked Schottky groups of genus $g$ is
parametrised by $3g-3$ coordinates
\be\label{Schcoord} \{p_1,\ldots,p_g,a_3,\ldots,a_g,r_2,\ldots,r_g\}\ ,\ee so that
$\Sch_g$ can be identified with a domain in
$\CC^{g}\times\hat\CC^{2g-3}$. More precisely, the condition that
there are $2g$ non-intersecting circles $C_{\pm i}$, $i=1,\ldots,g$,
satisfying \eqref{schcircle} implies
\be\label{Schboundary}
|p_i|<\min\Bigl\{\Bigl|\frac{(x_j-x_i)(x_k-x_{-i})}{(x_k-x_i)(x_j-x_{-i})}\Bigr|:
j,k=\pm 1,\ldots,\pm g\ ,j,k\neq \pm i\Bigr\}\ ,\quad i=1,\ldots,g\
, \ee
 where $x_i:=a_i$, $x_{-i}:=r_i$, $i=1,\ldots,g$. The space $\Sch_g$ is a
finite covering of the moduli space $\M_g$ of Riemann surfaces.
Given a Riemann surface $\Sigma$ uniformised by the Schottky group
$\Gamma$, we can take the homology classes of $C_1,\ldots,C_g$ as
the generators $\alpha_1,\ldots,\alpha_g$ in the symplectic basis of
$H_1(\Sigma,\ZZ)$ (see appendix \ref{a:Riemann}). Thus, the choice
of the group $\Gamma$ uniformising $\Sigma$ also gives a canonical
choice for the basis $\omega_1,\ldots,\omega_g$ of holomorphic
$1$-differentials satisfying \eqref{e:modnorm}. The Riemann period
matrix $\Omega$ depends also on the choice of the $\beta$-cycles.
However, the multiplicative periods $\exp (2\pi i\Omega_{ij})$ are
well-defined functions of the parameters \eqref{Schcoord}. More
precisely, one can prove that
\begin{align}\label{schtosiegela}
q_{ii}=e^{2\pi i\Omega_{ii}}=&p_i\prod_{\gamma\in \langle
\gamma_i\rangle \backslash \Gamma /\langle \gamma_i\rangle }\frac{(a_i-\gamma(a_i))(r_i-\gamma(r_i))}{(a_i-\gamma(r_i))(r_i-\gamma(a_i))}\\
q_{ij}=e^{2\pi i\Omega_{ij}}=&\prod_{\gamma\in \langle
\gamma_i\rangle \backslash \Gamma /\langle \gamma_j\rangle}
\frac{(a_i-\gamma(a_j))(r_i-\gamma(r_j))}{(a_i-\gamma(r_j))(r_i-\gamma(a_j))}\
,\qquad i\neq j\ ,\label{schtosiegelb}
\end{align}
where $\langle \gamma_i\rangle$ is the group generated by
$\gamma_i$, $i=1,\ldots,g$. Thus, for each element in $\langle
\gamma_i\rangle \backslash \Gamma /\langle \gamma_j\rangle$, we can
choose a representative with reduced word
$\gamma_{k_1}\cdots\gamma_{k_n}$, for $k_1\neq i$ and $k_n\neq j$.
\bigskip

\noindent
In the case of $g=2$, the Schottky space $\Sch_2$ is parametrised by
\be p_1\ ,\qquad p_2\ ,\qquad r_2\equiv x\ ,
\ee
and the condition \eqref{Schboundary} gives the identification
\be\label{schspace} \Sch_2\equiv\Bigl\{(p_1,p_2,x)\in\CC\times\CC\times\hat\CC\,:\,x\neq
0,1,\infty,\quad 0<|p_i|<\min\Bigl(|x|,\frac{1}{|x|}\Bigr),\,i=1,2\Bigr\}\
.\ee
The power series expansions of the multiplicative periods are
\begin{align}\label{q11exp}
q_{11}=e^{2\pi
i\Omega_{11}}=&p_1\sum_{n,m=0}^\infty\sum_{r=-n-m}^{n+m}
c(n,m,|r|)\, p_1^np_2^mx^r\ ,\\ \label{q22exp} q_{22}=e^{2\pi
i\Omega_{22}}=&p_2\sum_{n,m=0}^\infty\sum_{r=-n-m}^{n+m}
c(m,n,|r|)\, p_1^np_2^mx^r\ ,\\
\label{q12exp}q_{12}=e^{2\pi
i\Omega_{12}}=&x+x\sum_{n,m=1}^\infty\sum_{r=-n-m}^{n+m} d(n,m,r)\,
p_1^np_2^mx^r\ .
\end{align}
The coefficients $c(m,n,r)$ for $n\le 6$ and $m\le 7$ are listed in
tables \ref{t:schq1c} and \ref{t:schq1d} in appendix~\ref{a:tables},
while $d(m,n,r)=d(n,m,r)$ for $m,n\le 7$ are listed in tables
\ref{t:schzb} and \ref{t:schzc}.

\section{Partition functions}\label{a:partfunct}

The genus $g$ partition function $Z_g$ of a chiral conformal field
theory is defined, in physics, as the vacuum expectation value of
the theory on the  Riemann surface of genus $g$. Due to the
conformal anomaly, the partition function, however, depends not only
on the complex structure on the surface, but also on the specific
choice of coordinates on it. As a consequence, $Z_g$ cannot be
defined as a globally holomorphic function on $\M_g$, but rather
only as a holomorphic section on a suitable line bundle on $\M_g$.
(Alternatively, we may define $Z_g$ as a holomorphic function on
some covering space of $\M_g$, such as the space of Riemann period
matrices $\J_g\subseteq\Sieg_g$ or the Schottky space $\Sch_g$.)
More precisely, the partition function of a chiral conformal field
theory with central charge $c$ can be represented as a holomorphic
section\footnote{We observe that $\Li^{c/2}$ is a well-defined line
bundle on $\M_g$ only if $c$ is multiple of $4$, which is the case
for meromorphic conformal field theories. In the other cases, it can
only be defined as a \emph{projective} line bundle
\cite{Friedan:1986ua,Segal:2002ei}.} of $\Li^{\otimes c/2}$
\cite{Friedan:1986ua,Segal:2002ei}, where $\Li$ is the Hodge bundle.

The Hodge bundle $\Li$ can be described as follows. Consider the
vector bundle $\Lambda_g$ of rank $g$ on $\M_g$, whose fiber at the
point corresponding to the Riemann surface $\Sigma$ is the
$g$-dimensional vector space of holomorphic $1$-differentials on
$\Sigma$. As shown in appendix \ref{a:Riemann}, the choice of a
symplectic basis for the first homology group $H_1(\Sigma,\ZZ)$
determines a basis $\{\omega_1,\ldots,\omega_g\}$ of holomorphic
$1$-differentials on $\Sigma$, and hence a basis of local sections
on $\Lambda_g$, which we also denote by $\omega_1,\ldots,\omega_g$.
The line bundle $\Li$ is then defined as the $g$-th exterior product
of $\Lambda_g$, and given a choice of a basis for $H_1(\Sigma,\ZZ)$,
$\omega_1\wedge\ldots\wedge\omega_g$ defines a local holomorphic
section in $\Li$.
Under a symplectic transformation \eqref{modulhomol} the
corresponding local section of $\Li$ transforms as
\be
\omega_1\wedge\ldots\wedge\omega_g\ \mapsto\
\det(C\Omega+D)^{-1}(\omega_1\wedge\ldots\wedge\omega_g)\ ,\quad
\hbox{where} \quad
\left(\begin{matrix} A & B\\
C&D
\end{matrix}\right)\in\Sp(2g,\ZZ) \ . \ee
The partition function $Z_g$ of a meromorphic CFT is a global
holomorphic section of $\Li^{c/2}$, so that it can be written
locally as
$$Z_g=W_g(\Omega)\, (\omega_1\wedge\ldots\wedge\omega_g)^{c/2}\ ,
$$
where $W_g$ is a holomorphic function on the space $\J_g\subset
\Sieg_g$ of period matrices of Riemann surfaces. Since the section
cannot depend on the choice of the local trivialization, $W_g$ must
transform as a modular form of weight $c/2$
\be\label{modutW}W_g\Bigl((A\Omega+B)(C\Omega+D)^{-1}\Bigr) =\det(C\Omega+D)^{c/2}
\, W_g (\Omega)\ ,\ee under the action of $\left(\begin{smallmatrix}
A & B\\ C&D
\end{smallmatrix}\right)\in\Sp(2g,\ZZ)$.

\bigskip

Property \eqref{modutW} implies, in particular, that $W$ only
depends on the multiplicative periods $e^{2\pi i\Omega_{ij}}$ and
one can compose $W$ with \eqref{schtosiegela} and
\eqref{schtosiegelb} to define a function $\hat W$ on the Schottky
space. The Schottky uniformisation describes each Riemann surface as
a quotient of the Riemann sphere $\hat\CC$ by a discrete group, so
that the coordinates on $\hat\CC$ canonically define a system of
local complex coordinates on the Riemann surface. It is tempting to
conjecture that $\hat W$ is exactly the partition function $\hat
Z_g$ with respect to such coordinates. This is not true, and the
more general relation holds
\be
\hat Z_g=\frac{\hat W_g}{F_g^{c/2}}\ ,
\ee
for a certain holomorphic function $F_g$ on $\Sch_g$. It is clear
that the function $F_g$ does not depend on the theory in question, so
that it may be computed for some particular conformal field theory. For example, by
considering the conformal field theories associated to even
unimodular lattices, it is natural to conjecture that $F_g$ is the
function defined in \cite{McIntTak}, given by
\be
F_g=\prod_{m=1}^\infty\prod_{\gamma\neq 1}(1-p_{\gamma}^m)\ ,
\ee
where $p_\gamma$ is the multiplier of $\gamma$ and the product
runs over all the primitive classes in $\Gamma$, {\it i.e.}\ the
$PSL(2,\CC)$-conjugacy classes such that $\gamma$ is not conjugated
to any power $\tilde\gamma^n$, $n>1$. 
This infinite product converges on a certain open subset of $\Sch_g$,
and $F_g$ can be analytically continued to the whole $\Sch_g$ as a
holomorphic function. Similarly to the case of multiplicative
periods, the function $F_2$ can be written as a series in $p_1,p_2$.
For our purposes, we need the power $F_2^{-12}$, given by
\be\label{zogF}
F_2^{-12}=\sum_{m,n=0}^\infty \sum_{r=-m-n}^{m+n}b(m,n,|r|)\,
p_1^mp_2^nx^r \ , \ee where the coefficients $b(m,n,r)=b(n,m,r)$ for
$m,n\le 7$ are listed in tables \ref{t:schfua} and \ref{t:schfub} in
appendix~\ref{a:tables}.

\section{Partition function coefficients ad $2g$-point functions on the
sphere}\label{a:VinVout}


A holomorphic function on the Schottky space $\Sch_g$ can be
expanded in a power series in the multipliers $p_1,\ldots,p_g$
\be\label{finalZg}
\hat Z_g=\sum_{h_1,\ldots,h_g}p_1^{h_1}\cdots p_g^{h_g}\,
C_{h_1,\ldots,h_g}(a_3,\ldots,a_g,r_2,\ldots,r_g)\ , \ee where we
set $a_1=0$, $r_1=\infty$, $a_2=1$. If $\hat Z_g$ is the genus $g$
partition function of a consistent conformal field theory, the
functions $C_{h_1,\ldots,h_g}$ have a natural interpretation as
$2g$-point functions on the sphere. In this section, we will give a
heuristic justification of this relationship, following Segal's
approach to conformal field theories \cite{Segal:2002ei}.

In Segal's approach we can define amplitudes for any Riemann surface
with para\-me\-trised boundary circles $C_i$. The degrees of freedom of each boundary
circle are labelled by the vector space $\Hh$ of states; thus any such amplitude
also depends on the states $\phi_i\in\Hh$ that are associated to the boundary
circles $C_i$. (From a string theory perspective, the state $\phi_i$ describes
the external string state that is inserted at $C_i$.) Let us denote these amplitudes by
\be\label{segamp}
\Bigl\langle \prod_i \phi_i[C_i] \Bigr\rangle_{\cal D} \ ,
\ee
where ${\cal D}$ denotes a Riemann surface with boundary $\partial {\cal D} = \cup_i C_i$.

In the following we shall usually think of ${\cal D}$ as a subset of the complex plane.
We shall furthermore characterise the parametrised boundary circle $C_i$
by the M\"obius transformation $\gamma_i$ that maps the standard circle
\be
C := \{ z \in {\mathbb C} : |z| = 1 \}\ee to $C_i$, {\it i.e.}\ $C_i
= \gamma_i(C)$. Then we can identify the amplitude (\ref{segamp})
with
\be
\Bigl\langle \prod_i \phi_i[C_i] \Bigr\rangle_{\cal D}  =
\Bigl\langle \prod_i V\left(U(\gamma_i) \phi_i, \gamma_i(0) \right) \Bigr\rangle \ ,
\ee
where the amplitude on the right-hand-side is the standard amplitude in conformal
field theory, and
\be
U(\gamma) = \gamma'(0)^{L_0} \, e^{L_1
\frac{\gamma''(0)}{2\gamma'(0)}} \  . \ee

The advantage of Segal's approach is that it allows one to glue
Riemann surfaces together along boundary circles. Suppose ${\cal
D}^{(1)}$ and ${\cal D}^{(2)}$ are two Riemann surfaces with
boundary circles $C_i^{(1)}$ and $C_j^{(2)}$, respectively. By the
usual plumbing fixture construction we can then define a Riemann
surface ${\cal D}$ by identifying a parametrised boundary circle of
${\cal D}^{(1)}$, say $C_1^{(1)}=\gamma_{(1)}C$, with a parametrised
boundary circle of ${\cal D}^{(2)}$, say $C_1^{(2)}=\gamma_{(2)}C$.
%
This procedure is well defined provided that
\be
\gamma_{(1)}^{-1}\gamma_{(2)}:C\to \bar{C}
\ee
reverses the orientation of the standard circle. Here $\bar{C}$ denotes the
standard circle with the opposite orientation, so that $\bar{C}=\hat\gamma(C)$
with
\be
\hat\gamma (z) = -\frac{1}{z} \ . \ee If this is the case, then we
can identify the circles $C_1^{(1)}$ and $\bar
C_1^{(2)}=\gamma_{(2)}\hat\gamma C$ via
\be\label{identcirc}
\gamma_{(1)}\hat\gamma\gamma_{(2)}^{-1}:
\bar C_1^{(2)}\to C_1^{(1)}\ ,
\ee
and the maps $\gamma_{(1)}\hat\gamma$ and $\gamma_{(2)}$
define local analytic coordinates on a neighbourhood of
$C_1^{(1)}\equiv \bar C_1^{(2)}$ in $\D$.  The amplitude on the
${\cal D}$ is then simply
\begin{eqnarray}\label{segglue}
& & \Bigl\langle \prod_{i\geq 2} \phi_i[C_i^{(1)}]\, \prod_{j\geq 2} \phi_j[C_j^{(2)}]
\Bigr\rangle_{\cal D}  \nonumber \\
& & \qquad \qquad =
\sum_{\psi_1,\psi_2} G_{\psi_1,\psi_2}^{-1}
\Bigl\langle \psi_1[C_1^{(1)}]\,  \prod_{i\geq 2} \phi_i[C_i^{(1)}]\Bigr\rangle_{{\cal D}_1}
\Bigl\langle \psi_2[C_1^{(2)}]\, \prod_{j\geq 2} \phi_j[C_j^{(2)}]\Bigr\rangle_{{\cal D}_2} \ ,
\end{eqnarray}
where
\be
G_{\phi\psi}=\lim_{z\to \infty}\langle V(z^{2L_0}e^{zL_1}\phi,z)V(\psi,0)\rangle
\ee
is the metric on the space of states (with $G_{\phi\psi}^{-1}$ the inverse metric).\footnote{
Note that our metric differs from the standard Zamolodchikov metric $G^{Z}$ by a sign,
$G_{\phi\psi}= G^{Z}_{(-1)^{L_0}\phi,\psi}$.}
\medskip

\noindent
To illustrate this general construction, consider the annulus
\be
\A:=\{z\in\CC: |q|<|z|<1\}
\ee
for some $q\in\CC$, $0<|q|<1$. The boundary components are
$\bar C=\hat\gamma(C)$ and $\gamma_q(C)$, where
$\gamma_q(z)= qz$. The amplitudes on $\A$ are defined as
\be \bigl\langle  \phi[\hat\gamma C]\;\psi[\gamma_qC] \bigr\rangle_\A  =
\lim_{z\to \infty}\bigl\langle
V(z^{2L_0}e^{zL_1}\phi,z)V(q^{L_0}\psi,0)\bigr\rangle\ .
\ee
Using the prescription \eqref{segglue} to glue the boundary circles of the annulus
to form the torus with modular parameter $q$ then leads to
\be \sum_{\phi,\psi}G_{\phi,\psi}^{-1}
\Bigl\langle  \phi[\gamma_qC]\phi[\hat\gamma C]
\Bigr\rangle_\A=
\sum_{\phi,\psi}G_{\phi,\psi}^{-1}
\lim_{z\to \infty}\left\langle
V\Bigl(z^{2L_0}e^{zL_1}\phi,z\Bigr)\,
V\bigl(q^{L_0}\psi,0\bigr)\right\rangle=\Tr_{\Hh}(q^{L_0}) \ ,
\ee
which is indeed the expected result.
\bigskip

Let us now consider the case of a Riemann surface of genus $g$,
uniformised by a Schottky group $\Gamma$ with generators
$\gamma_1,\ldots,\gamma_g$, and let $p_i$, $a_i$ and $r_i$ be
the multiplier, and the attractive and repelling fixed points of
$\gamma_i$, respectively. Let us define the M\"obius transformations
\be\label{moeb} \gamma_{a_i,r_i}(z)=\frac{r_iz+a_i}{z+1}\ ,
\ee
satisfying $\gamma_{a_i,r_i}(0)=a_i$ and $\gamma_{a_i,r_i}(\infty)=r_i$, so that
the generators of $\Gamma$ can be written as
\be\label{losgeneratores}
\gamma_i=\gamma_{a_i,r_i}\gamma_{p_i}\gamma_{a_i,r_i}^{-1}\ ,\qquad i=1,\ldots,g\ ,
\ee
where $\gamma_{p_i}$ is defined as before by $\gamma_p(z)=pz$. A fundamental
domain for $\Gamma$ is given by
\be\D:=\hat\CC\setminus\bigcup_{i=1}^g(D_i\cup D_{-i})\ ,\ee
where
\begin{align}
D_i&=\Bigl\{z\in\CC:\frac{|z-a_i|}{|z-r_i|}<|R_i|\Bigr\}=\gamma_{a_ir_i}\gamma_{R_i}(D)\ ,\\
D_{-i}&=\Bigl\{z\in\CC:\frac{|z-r_i|}{|z-a_i|}<|R_{-i}|\Bigr\}=\gamma_{a_ir_i}\hat\gamma\gamma_{R_{-i}}(D)\ ,
\end{align}
and $D$ is the unit disc,
\be D=\{z\in\CC:|z|<1\}\ .
\ee Here $R_i,R_{-i}\in\CC$ are chosen such that
\be\label{riripi}
R_{-i}R_i=p_i\ ,
\ee
and all disks $D_i,D_{-i}$, $i=1,\ldots,g$ are disjoint. (Such $R_i,R_{-i}$ exist only if
eq.~\eqref{Schboundary} is satisfied). The boundary
$\partial\D=\bigcup_i C_{i}\cup \bar C_{-i}$ has
components
\be C_{i}=\gamma_{a_i,r_i}\gamma_{R_i} C\ ,\qquad
\bar C_{-i}=\gamma_{a_i,r_i}\hat\gamma\gamma_{R_{-i}}C\ .
\ee
We note that
\be C_{-i}\equiv \bar{\bar{C}}_{-i}=\gamma_{a_i,r_i}\hat\gamma
\gamma_{R_{-i}}\hat\gamma\, C=\gamma_{a_i,r_i}\gamma_{R_{-i}}^{-1}C\
, \ee since
$\hat\gamma\gamma_{R_{-i}}\hat\gamma=\gamma_{1/R_{-i}}=\gamma_{R_{-i}}^{-1}$.
The Riemann surface can be obtained by gluing each $C_i\equiv
C_{-i}$ according to the general procedure described above. In this
case, because of \eqref{losgeneratores} and \eqref{riripi}, the
identification map \eqref{identcirc} is simply $\gamma_i:C_{-i}\to
C_i$, in agreement with \eqref{schcircle}. Using the gluing
prescription \eqref{segglue}, the partition function $\hat Z_g$ is
then
\be
\begin{split}
\hat Z_g&=\sum_{\phi_i\psi_i\in\Hh}\prod_iG^{-1}_{\phi_i\psi_i}
\Bigl\langle\prod_i\phi_i[C_i]\psi_i[\bar C_{-i}]\Bigr\rangle_{\D}\\
&=\sum_{\phi_i\psi_i\in\Hh}\prod_iG^{-1}_{\phi_i\psi_i}\Bigl\langle\prod_i
V\Bigl(U(\gamma_{a_i,r_i})R_i^{L_0}\phi_i,a_i\Bigr)
\;V\Bigl(U(\gamma_{a_i,r_i}\hat\gamma)R_{-i}^{L_0}\psi_i,r_i\Bigr)\Bigr\rangle\ .
\end{split} \ee
If we choose $\phi_i,\psi_i$ to be eigenvectors of $L_0$
and use \eqref{riripi}, we finally obtain \eqref{finalZg}, with
\be\label{genC2gpt}
C_{h_1,\ldots,h_g}
=\sum_{\phi_i,\psi_i\in\Hh_{h_i}} \prod_{i=1}^g
G_{\phi_i\psi_i}^{-1}\Bigl\langle\prod_{i=1}^g
V^{in}(\phi_i,a_i)\;V^{out}(\psi_i,r_i)\Bigr\rangle\ ,\ee where
\begin{align}\label{e:Vin}
V^{in}(\phi,a_i)&=
V\bigl(U(\gamma_{a_i,r_i})\phi,a_i\bigr)
=V\Bigl((r_i-a_i)^{L_0}e^{-L_1}\phi,a_i\Bigr)\\
\label{e:Vout}V^{out}(\psi,r_i)&=
V\bigl(U(\gamma_{a_i,r_i}\hat\gamma)\psi,r_i\bigr)
=V\Bigl((r_i-a_i)^{L_0}e^{L_1}\psi,r_i\Bigr)\ .
\end{align}
Note that $\hat Z_g$ is independent of the specific choice of
$R_i,R_{-i}$ satisfying \eqref{riripi}. Equivalently,
$C_{h_1,\ldots,h_g}$ is not affected by any replacement
$\gamma_{a_i,r_i}\to\gamma_{a_i,r_i}\gamma_t$ with $t\in\CC^*$, in
the definition of $V^{in}$ and $V^{out}$. When all states in
$\Hh_{h_1},\ldots,\Hh_{h_g}$ are quasi-primaries, \eqref{genC2gpt}
simplifies to
\be\label{C2gpoint}
\C_{h_1,\ldots,h_g}=
\prod_{i=1}^g(r_i-a_i)^{2h_i}
\sum_{\phi_i,\psi_i\in\Hh_{h_i}}\prod_{i=1}^gG^{-1}_{\phi_i\psi_i}
\Bigl\langle\prod_{i=1}^gV(\phi_i,r_i)V(\psi_i,a_i)\Bigr\rangle\ .
\ee
\bigskip

\noindent
At genus $g=2$, eq.~(\ref{genC2gpt}) can be written as
\be C_{h_1,h_2}(x)=\sum_{\phi_i,\psi_i\in\Hh_{h_i}}
G_{\phi_1\psi_1}^{-1}G_{\phi_2\psi_2}^{-1}\Bigl\langle
V^{out}(\psi_1,\infty)\;V^{out}(\psi_2,x)\;V^{in}(\phi_2,1)\;V^{in}(\phi_1,0)\Bigr\rangle\ ,
\ee
where
\begin{align}
 V^{out}(\psi_1,\infty)&=V(U(\hat\gamma)\psi_1,\hat\gamma(0))\qquad
& V^{in}(\phi_1,0)&=V(\phi_1,0)\\
V^{out}(\psi_2,x)&=V(U(\gamma_{1,x}\hat\gamma)\psi_2,x)\qquad
&V^{in}(\phi_2,1)&=V(U(\gamma_{1,x})\phi_2,1)\ .
\end{align}
As mentioned before, we can equivalently replace $\gamma_{1,x}$ by
any $\gamma_{1,x}\gamma_t$ for $t\in\CC^*$. In particular, if we
replace $\gamma_{1,x}$ by the involution
\be
\tilde\gamma_{1,x}(z)=\frac{z-1}{z/x-1}=\gamma_{1,x}\gamma_{-1/x}\ ,
\ee
the symmetries of $C_{h_1,h_2}(x)$ are immediate.
Indeed, using the invariance of the $4$-point amplitude under the
M\"obius transformation $\tilde\gamma_{1,x}$, we have
\be\label{symmh1h2}\begin{split}
C_{h_1,h_2}(x)
=&\sum_{\phi_i,\psi_i\in\Hh_{h_i}}
G_{\phi_1\psi_1}^{-1}G_{\phi_2\psi_2}^{-1}\Bigl\langle
V(U(\tilde\gamma_{1,x}\hat\gamma)\psi_1,x)V(U(\hat\gamma)\psi_2,\infty)
V(\phi_2,0)V(U(\tilde\gamma_{1,x})\phi_1,1)\Bigr\rangle\\
=&\sum_{\phi_i,\psi_i\in\Hh_{h_i}}
G_{\phi_1\psi_1}^{-1}G_{\phi_2\psi_2}^{-1}\Bigl\langle
V^{out}(\psi_1,x)\;V^{out}(\psi_2,\infty)\;V^{in}(\phi_2,0)\;V^{in}(\phi_1,1)\Bigr\rangle \\
=& C_{h_2,h_1}(x)\ .
\end{split}
\ee
Similarly, applying the M\"obius transformation
$\hat\gamma$ and noting that
$\hat\gamma\tilde\gamma_{1,x}\hat\gamma=\tilde\gamma_{1/x,1}$, we
obtain
\begin{eqnarray}\label{symm1x}
& & \hspace*{-0.2cm} C_{h_1,h_2}(x) \nonumber \\
& & = \sum_{\phi_i,\psi_i\in\Hh_{h_i}}
G_{\phi_1\psi_1}^{-1}G_{\phi_2\psi_2}^{-1}\Bigl\langle
V(\psi_1,0)V(U(\tilde\gamma_{1/x,1})\psi_2,1/x)
V(U(\tilde\gamma_{1/x,1}\hat\gamma)\phi_2,1)V(U(\hat\gamma)\phi_1,\infty)\Bigr\rangle
\nonumber \\
&& = \sum_{\phi_i,\psi_i\in\Hh_{h_i}}
G_{\phi_1\psi_1}^{-1}G_{\phi_2\psi_2}^{-1}\Bigl\langle
V^{in}(\psi_1,0)\;V^{in}(\psi_2,1/x)\;V^{out}(\phi_2,1)\;V^{out}(\phi_1,\infty)\Bigr\rangle
\nonumber \\
& & =C_{h_1,h_2}(1/x)\ .
\end{eqnarray}
The same argument also applies to the quasiprimary functions, defined in
\eqref{fourpprim}.

\section{Technicalities}\label{app:CD}
\subsection{The relation between $\D$ and $\C$}

In this appendix we will prove, using the associativity of the OPE, that
one can
always express $\C^{(*)}_{h_1,h_2;l}$ in terms of
$\D_{h_1',h_2';l'}$ with $h_1'\le h_1$, $h_2'\le h_2$ and $l'\le l$
and vice versa. In fact we will prove that the linear spans of these
coefficients are related as
\be\label{D1}
\bigl\langle
\C^{(*)}_{h_1,h_2';l}\bigr\rangle_{\substack{l=0,\ldots,L\\h_2'\le
h_2}}=\bigl\langle
\D_{h_1,h_2';l}\bigr\rangle_{\substack{l=0,\ldots,L\\h_2'\le h_2}} \
. \ee To do this it is useful to introduce yet another set of
invariants $\C^{(1)}_{h_1,h_2;l}$ as
\be
\C_{h_1,h_2}(x) = \sum_{l=0}^\infty (x-1)^l \C^{(1)}_{h_1,h_2;l}\ .
\ee
Using the same arguments as in section~3.1.1 it follows that
these invariants can be interpreted as
\be \label{c1}
\C^{(1)}_{h_1,h_2;l}:=\sum_{\phi_2,\psi_2\in\Hh_{h_2}^{qp}}
G^{-1}_{\phi_2\psi_2}\Tr_{\Hh_{h_1}}\bigl(V_0(V_{h_2-l}(\phi_2)\psi_2)\bigr)\
. \ee In order to apply these results also to the case of section
\ref{s:PFconstr}, we will not assume that $\Hh$ is a consistent self
dual conformal field theory, but just the representation of a chiral
algebra $\A$, so that \eqref{ansatz} holds for some $B>0$ (that is,
all the fields up to weight $B$ are in the vacuum representation
$\Hh^{(0)}$ of a consistent chiral algebra). Notice that \eqref{c1}
still makes sense for $h_2<B$, because
$V_{h_2-l}(\phi_2)\psi_2\in\Hh^{(0)}$, so we will restrict ourselves
to this case.

\noindent
Using a similar argument as in (\ref{2.32}) one shows that
\be \label{idCstarC1}
\langle \C^{(*)}_{h_1,h_2;l} \rangle_{l=0,\ldots,L} = \langle
\C^{(1)}_{h_1,h_2;2l}\rangle_{l=0,\ldots,L}\ . \ee Note the
appearance of $2l$ for the $\C^{(1)}_{h_1,h_2;l}$, which comes from
the fact that the leading power in the term $\C^{(*)}_{h_1,h_2;l}$
is $(x-1)^{2l}$. We now want to show that the invariants
$\C^{(1)}_{h_1,h_2;2l}$
can be written in terms of the invariants $\D_{h_1,h_2;l}$. Let
$\phi_2,\psi_2\in\Hh^{(0)}_{h_2}$ be quasiprimary states and
consider
\begin{align} \label{completeset1}
V_0\bigl(V_{h_2-2l}(\phi_2)\psi_2\bigr)=&
\sum_{\phi,\psi\in\Hh_{2l}\cap\Hh^{(0)}}G^{-1}_{\phi\psi}\,
V_0(\phi)\, 
\bigl\langle\psi|V_{h_2-2l}(\phi_2)\psi_2\bigr\rangle\ ,
\end{align} where $\langle\phi|\psi\rangle\equiv G_{\phi\psi}$.
The sum on the right hand side can be taken over states of the form
$\psi=L_{-1}^n\psi'$ and $\phi=L_{-1}^m\phi'$, with $m,n\ge 0$ and
$\phi'$ and $\psi'$ quasiprimary states of weight $2l-n$ and $2l-m$.
Note that
\be G_{\phi\psi}=\bigl\langle\phi'|L_1^mL_{-1}^n\psi'\bigr\rangle=\delta_{mn}n!(4l-2n)\cdots(4l-n-1)G_{\phi'\psi'}\
, \ee
\be\label{V0desc} V_0(\phi)=V_0(L_{-1}^m\phi')=(-1)^m
(2l-1)(2l-2)\cdots(2l-m)V_0(\phi')\ ,\ee and
\begin{eqnarray}
\bigl\langle\psi|V_{h_2-2l}(\phi_2)\psi_2\bigr\rangle & = &
\bigl\langle\psi'|L_1^{n}V_{h_2-2l}(\phi_2)\psi_2\bigr\rangle  \\
& = &
(2l-1)(2l-2)\cdots(2l-n)\bigl\langle\psi'|V_{h_2-(2l-n)}(\phi_2)\psi_2\bigr\rangle
\ . \nonumber
\end{eqnarray}
It thus follows that $V_0(V_{h_2-2l}(\phi_2)\psi_2)$ can be
expressed as a linear combination of
\be \sum_{\phi,\psi\in\Hh_{l'}^{qp}\cap\Hh^{(0)}}
G^{-1}_{\phi\psi}V_0(\phi)\bigl\langle\psi|V_{h_2-l'}(\phi_2)\psi_2\bigr\rangle=
\sum_{\phi,\psi\in\Hh_{l'}^{qp}\cap\Hh^{(0)}}
G^{-1}_{\phi\psi}V_0(\phi)\lim_{z\to\infty}z^{2l'}\langle
\psi(z)\phi_2(1)\psi_2(0)\rangle\ee with $l'=0,\ldots,2l$. Applying
a fractional linear transformation to the last correlator, we can
exchange $1$ and $\infty$ while keeping $0$ fixed. Because all the
states in this correlator are quasiprimary, we have simply
\be\lim_{z\to\infty}z^{2l'}\langle
\psi(z)\phi_2(1)\psi_2(0)\rangle=(-1)^{l'}\lim_{\zeta\to\infty}\zeta^{2h_2}
\langle
\phi_2(\zeta)\psi(1)\psi_2(0)\rangle=(-1)^{l'}\bigl\langle\phi_2|V_0(\psi)\psi_2\bigr\rangle\
. \ee
%
%
%
Thus
$\C^{(1)}_{h_1,h_2;2l}$, $h_2<B$ is a linear combination of
\begin{eqnarray}\label{totorus}
& & \sum_{\substack{\phi_2,\psi_2\in\Hh_{h_2}^{qp}\\
\phi,\psi\in\Hh_{2l'}^{qp}\cap\Hh^{(0)}}}
G^{-1}_{\phi\psi}G^{-1}_{\phi_2\psi_2}\Tr_{\Hh_{h_1}}\bigl(V_0(\phi)\bigr)
\bigl\langle\phi_2|V_0(\psi)\psi_2\bigr\rangle \nonumber \\
& & \qquad \qquad \qquad
=\sum_{\phi,\psi\in\Hh_{2l'}^{qp}\cap\Hh^{(0)}} G^{-1}_{\phi\psi}\,
\Tr_{\Hh_{h_1}}\bigl(V_0(\phi)\bigr)
\Tr_{\Hh_{h_2}^{qp}}\bigl(V_0(\psi)\bigr)
\end{eqnarray}
for $l'=0,\ldots,l$, and therefore, by \eqref{V0desc}, it is
also a linear combination of
\be \sum_{\phi,\psi\in\Hh_{2l'}\cap\Hh^{(0)}}
G^{-1}_{\phi\psi}\,
\Tr_{\Hh_{h_1}}\bigl(V_0(\phi)\bigr)\Tr_{\Hh_{h_2}^{qp}}\bigl(V_0(\psi)\bigr)\
,\qquad l'=0,\ldots,2l\ . \ee Next we observe that for any
$\psi\in\Hh^{(0)}$, we have
\be
\Tr_{\Hh_{h_2}}\bigl(V_0(\psi)\bigr)=
\sum_{n=0}^{h_2}\Tr_{L_{-1}^n\Hh_{h_2-n}^{qp}}\bigl(V_0(\psi)\bigr)
=
\sum_{n=0}^{h_2}c(h_2,n)\Tr_{\Hh_{h_2-n}^{qp}}\bigl(V_0(\psi)\bigr)
\ , \ee for some coefficients $c(h_2,n)$. These identities can be
inverted to obtain $\Tr_{\Hh_{h_2}^{qp}}\bigl(V_0(\psi)\bigr)$ as a
linear combination of $\Tr_{\Hh_{h_2'}}\bigl(V_0(\psi)\bigr)$ with
$h_2'\le h_2$. Furthermore, by \eqref{e:L0def}, for all $N\ge 0$ we
have
\be \bigoplus_{n=0}^N \Hh_n=\bigoplus_{n=0}^N \Hh_{[n]}\ .
\ee
Thus, by \eqref{totorus}, $\C^{(1)}_{h_1,h_2;2l}$ is a linear
combination of $\D_{h_1,h_2';l'}$ with $l'\le l$ and $h_2'\le h_2$.
More precisely, for any $h_2\le h_1$, $h_2\le B$ and $L\ge 0$, we
have
\be \bigl\langle
\C^{(1)}_{h_1,h_2';2l}\bigr\rangle_{\substack{l=0,\ldots,L\\h_2'\le
h_2}}=\bigl\langle
\D_{h_1,h_2';l}\bigr\rangle_{\substack{l=0,\ldots,L\\h_2'\le h_2}}\
.\ee Together with (\ref{idCstarC1}) this then immediately implies
(\ref{D1}).

\subsection{Completeness of the relations}\label{app:Thm1}
In this appendix we prove Theorem~\ref{ThmComplete}. For what
follows it will be useful to introduce a lexicographical ordering on
the space of coefficients. We say that
\be
(h_1,h_2;l) \prec (\hat{h}_1, \hat{h}_2; \hat{l}) \ee if (1)
$h_2<\hat{h}_2$, or (2) $h_2=\hat{h}_2$ and $h_1<\hat{h}_1$, or (3)
$h_2=\hat{h}_2$, $h_1=\hat{h}_1$ and $l < \hat{l}$. We define the
relation $\preceq$ in the obvious way.

To prove Theorem~\ref{ThmComplete}, first note that by \eqref{rel1}
and \eqref{rel2} we can express all invariants
$\C^{(0)}_{h_1,h_2;l}$ and $\D_{h_1,h_2;l}$ as linear combinations
of $\C^{(*)}_{h_1,h_2;l}$ with $l=0,\ldots, h_1+h_2$. We then want
to prove the following lemma:
\smallskip

\setcounter{theorem}{0}
\begin{lemma}\label{lemma1}
Let $(h_1,h_2;l) \notin {\cal P}^{(*)}_k$. It is then possible to
express $\C^{(*)}_{h_1,h_2;l}$ in terms of invariants
$\C^{(*)}_{h_1',h_2';l'}$ with $(h_1',h_2';l') \prec (h_1,h_2;l)$.
\end{lemma}
\smallskip

\noindent
By  repeatedly applying Lemma~\ref{lemma1} it is clear that we can express
any $\C^{(*)}$ invariant in terms of the invariants  $\C^{(*)}_{h_1,h_2;l}$
with $(h_1,h_2;l)\in{\cal P}^{(*)}_k$. From this Theorem~\ref{ThmComplete} follows.
\bigskip

\noindent {\bf Proof of Lemma 1.}
For $(h_1,h_2;l) \notin {\cal P}^{(*)}_k$ it is enough to treat two
different cases:
\begin{enumerate}
\item $l>h_2$:
It follows from (\ref{rel1}) that we can express
$\C^{(*)}_{h_1,h_2;l}$ as a linear combination of
$\C^{(0)}_{h_1,h_2;l'}$ with $l' = 0,1,\ldots ,L$ with $L =
h_1+h_2-l< h_1$. We can then use (\ref{rel3}) to rewrite
$\C^{(0)}_{h_1,h_2;l'} = \C^{(0)}_{l',h_2;h_1}$.
Using \eqref{rel1} again, these can in turn be expressed in terms of
$\C^{(*)}_{l',h_2;l''}$, where $(l',h_2;l'')\prec (h_1,h_2;l)$
because $l'<h_1$.

\item $h_1>k+l-w_l$ or $h_2>h_1$:
Using (\ref{rel2}) we can express $\C^{(*)}_{h_1,h_2;l}$ in terms of
$\D_{h_1',h_2';l'}$ with $(h_1',h_2';l') \preceq (h_1,h_2;l)$. Since
we can use (\ref{rel2}) to rewrite the $\D_{h_1',h_2';l'}$ in terms
of $\C^{(*)}_{h_1'',h_2'';l''}$ with $(h_1'',h_2'';l'') \preceq
(h_1',h_2';l')$, it is clear that we only need to worry about the
terms with $(h_1',h_2';l') = (h_1,h_2;l)$. In the case $h_2>h_1$, we
have the obvious relation $\D_{h_1,h_2;l}=\D_{h_2,h_1;l}$ and
$(h_2,h_1;l)\prec (h_1,h_2;l)$. In the case $h_1>k+l-w_l$, we can
use \eqref{rel4} to express $\D_{h_1,h_2;l}$ as a linear combination
of $\D_{h_1',h_2';l}$ with $h_2'\leq h_1' \leq k+l-w_l$ and $h_2'
\leq h_2$. From this we see that $h_1'< h_1$, so that after a
reconversion to $\C^{(*)}$, using (\ref{rel2}), the claim of the lemma
also follows in this case.
\end{enumerate}

\subsection{Free parameters and the slope of effective
divisors}\label{a:slope}

Recall that for $g=2$, the triples $(h_1,h_2,l)\in {\cal P}^{(*)}_k$ satisfy the
bound (see (\ref{2.29}))
\be\label{upperfree} h_1,h_2\le \frac{6}{5}k\ .
\ee In this section we will discuss a similar bound for ${\cal
P}_k(g)$, for general $g$. Our approach is similar to the procedure
adopted in \cite{Moore:1986sk} in the framework of string theory.
\smallskip

Recall from appendix~\ref{a:partfunct} that the genus $g$ partition
function of a chiral conformal field theory of central charge
$c=24k$ is a section of the tensor power $\Li^{\otimes 12k}$ of the
Hodge bundle on $\M_g$, whose fiber at a certain point $\Sigma$ is
$\bigwedge^g H^0(\Sigma,K_\Sigma)$. This line bundle naturally
extends to the Deligne-Mumford compactification
\be \bar\M_g=\M_g\cup\bigcup_{i=0}^{[g/2]}\Delta_i
\ee
of the moduli space. Here, a generic element in the boundary
component $\Delta_i$, $i>0$, is obtained by identifying a point on a
curve of genus $i$ with a point on a curve of genus $g-i$; a generic
element of $\Delta_0$ is obtained by identifying two distinct points
on a curve of genus $g-1$.

Let $Pic(\bar\M_g)$ be the group of (isomorphism classes of)
holomorphic line bundles on $\bar\M_g$, equipped with a tensor product
and with the inverse given by the dual line bundle. The Picard group
$Pic(\bar\M_g)$ is isomorphic\footnote{More precisely,
$Pic(\bar\M_g)\otimes\QQ$ is isomorphic to the group of rational
divisor classes on $\hat\M_g$ \cite{HarrMorr2}.} to the group of
divisor classes on $\bar\M_g$. Thus we shall from now on
adopt an additive notation for this group. Let us
define by\footnote{All the singular curves in $\Delta_1$ have a
non-trivial automorphism, acting as the involution of the torus with
one puncture. For this reason, it is convenient to include a factor
$1/2$ in the definition of $\delta_1$.}
\be
\delta_i=[\Delta_i]\ , \quad i\neq 1\ ,\qquad\quad
\delta_1=\frac{1}{2}[\Delta_1]
\ee
the divisor classes of the boundary components. It can be shown that
$Pic(\bar\M_g)$ is generated by
\be \lambda,\ \delta_0,\ \ldots,\ \delta_{[\frac{g}{2}]}\ ,
\ee where $\lambda$ denotes the divisor class of the Hodge bundle
$\Li$. (For $g>2$ there are no relations, whereas for $g=2$ there is
one relation, namely $10\lambda=\delta_0+2\delta_1$.)

It is clear that the elements of the set ${\cal P}_k(g)$ correspond
to a basis of the space of holomorphic sections of $\Li^{\otimes
12k}$. Let $Z$ and $Z'$ be two distinct sections of $\Li^{\otimes
12k}$. The divisor $(Z-Z')$ of their difference can be written as
\be\label{divisori}
(Z-Z')=D+b_0\,\Delta_0+\frac{1}{2}b_1\,\Delta_1+\sum_{i=2}^{[g/2]}b_i\,\Delta_i\
, \ee where $D$ is (the closure of) an effective divisor in $\M_g$,
and $b_i\ge 0$. Since $Z-Z'$ is again a section of $\Li^{\otimes
12k}$, eq.~\eqref{divisori} implies
\be [D]=12k\lambda-\sum_{i=0}^{[g/2]}b_i\delta_i\ ,\ee in terms of
divisor classes. Suppose that $Z-Z'$ vanishes at order at least
$L\ge 0$ in any degeneration limit, that is
\be\label{minbi} L=\min_i b_i\ .\ee
If we consider any expansion of $Z$ and $Z'$ of the form
\eqref{gengexp}, this means that
\be C_{h_1,\ldots,h_{3g-3}}=C'_{h_1,\ldots,h_{3g-3}}\ ,
\ee whenever $h_i<L$ for some $i$. Thus, two distinct $Z$ and $Z'$
satisfying \eqref{minbi} exist if and only if there is at least one
element in ${\cal P}_k(g)$ with $h_i\ge L$ for all $i$.
\smallskip

\noindent Following \cite{HarrMorr} we defined the slope $s_g$ as
\be
s_g=\inf\bigl\{\frac{a}{\min_i b_i}\mid a,b_i>0\, ,\ \exists \,
\text{effective divisor $D$ such that }
[D]=a\lambda-\sum_{i=0}^{[g/2]}b_i\delta_i\bigr\} \ .
\ee
Then $Z\neq Z'$ implies
\be \label{D.26}
L\le \frac{12k}{s_g}\ ,
\ee
and hence there is no element in ${\cal P}_k(g)$ with
$h_i>\frac{12k}{s_g}$ for all $h_i$.

For small values of $g$, the value of $s_g$ has been determined in
\cite{HarrMorr,Tan,FaPopa}. A lower bound valid for all $g$ (but in general not
sharp) is \cite{Pandha}
\be s_g\ge \frac{60}{g+4}\ .
\ee
Together with (\ref{D.26}) this then implies that
${\cal P}_k(g)$ does not contain any elements with
\be
h_1,\ldots,h_{3g-3}> \frac{k}{5}(g+4)\ ,
\ee
which is the desired inequality.

\subsection{Proof of formula \eqref{numstates}}\label{a:cardy}

The function $Z_1(q)$ is a modular form of weight $12k$, so that
\be \frac{Z_1(q)}{\Delta^{k}}=\sum_{h=0}^\infty n_hq^{h-k}
\ee
is modular invariant. This implies that it can be written as
\be\label{poneHecke} \frac{Z_1(q)}{\Delta^{k}}=n_k+\sum_{t=1}^k n_{k-t} T'_tJ(\tau)\ ,
\ee
where
\be J(\tau)=j(\tau)-744=\sum_{m=-1}^\infty c_mq^m
=q^{-1}+196884q+\ldots
\ee
is the Klein invariant and $T'_t$ is the Hecke operator
\be T'_tF(\tau)=\sum_{d|t}\sum_{b=0}^{d-1}F\Bigl(\frac{t\tau+bd}{d^2}\Bigr)\ .
\ee
If $F$ is modular invariant, then so is $T'_tF$. Furthermore,
\be\begin{split} T'_tJ(\tau)&=\sum_{d|t}\sum_{m=-1}^\infty
c_me^{2\pi im\frac{t\tau}{d^2}}\sum_{b=0}^{d-1}
e^{2\pi i\frac{mb}{d}}=\sum_{m=-1}^\infty c_m\sum_{d|(t,|m|)}d\,
e^{2\pi im\frac{t\tau}{d^2}}\\&=q^{-t}+\sum_{l=1}^\infty
q^l\sum_{d|(t,l)}\frac{t}{d}\;c_{tl/d^2}\ ,
\end{split}
\ee
where $(a,b)$ denotes the greatest common divisor.
Eq.~\eqref{poneHecke} follows immediately from these properties of
Hecke operators. Using Rademacher's formula, one has the following
asymptotic estimate for the coefficients of $J(\tau)$
\be c_m\sim \frac{m^{-3/4}}{\sqrt 2}\;e^{4\pi\sqrt{m}}\ ,\qquad m\gg0\
. \ee Thus, the coefficients $n_{l+k}$ of $Z_1/\Delta^{12k}$ for
large $l$ are given by
\be\begin{split} n_{l+k}&=\sum_{t=1}^k n_{k-t}\sum_{d|(t,l)}\frac{t}{d}\;c_{tl/d^2}\sim
\sum_{t=1}^k
n_{k-t}\sum_{d|(t,l)}\frac{t}{d}\;\frac{d^{3/2}}{\sqrt{2}(tl)^{3/4}}\;e^{4\pi\frac{\sqrt{tl}}{d}}\\
&\sim n_0\frac{k^{1/4}}{\sqrt{2}}\;l^{-3/4}\;e^{4\pi\sqrt{kl}}\ ,
\end{split} \ee where in the last step we retained only the leading terms $t=k$ and
$d=1$. This then implies eq.~\eqref{numstates} because $n_0=1$ follows from
the uniqueness of the vacuum.

\newpage \section{Tables}\label{a:tables}\nopagebreak[4] 
\nopagebreak[4]{\scriptsize\fontsize{10pt}{12pt}\selectfont
\begin{align*}
&\begin{array}{|c|rrrrrrrr|}\hline c(m,n,|r|) &
\multicolumn{8}{|c|}{m,n}\\
\hline |r| & 0,0 & 0,1 & 0,2 & 0,3 & 0,4 & 0,5 & 0,6 &0,7 \\
\hline
 0 & 1 & -4 & 6 & 0 & -8 & 0 & 10 & 0 \\
 1 & 0 & 2 & -6 & 4 & 4 & 0 & -10 & 0 \\
 2 & 0 & 0 & 3 & -8 & 5 & 0 & 5 & 0 \\
 3 & 0 & 0 & 0 & 4 & -10 & 6 & 0 & 0 \\
 4 & 0 & 0 & 0 & 0 & 5 & -12 & 7 & 0 \\
 5 & 0 & 0 & 0 & 0 & 0 & 6 & -14 & 8 \\
 6 & 0 & 0 & 0 & 0 & 0 & 0 & 7 & -16 \\
 7 & 0 & 0 & 0 & 0 & 0 & 0 & 0 & 8
\\ \hline
\end{array}\\&
\begin{array}{|c|rrrrrrrr|}\hline c(m,n,|r|) &
\multicolumn{8}{|c|}{m,n}\\ \hline
|r|  & 1,0 & 1,1 & 1,2 & 1,3 & 1,4 & 1,5 & 1,6 &1,7 \\
\hline
 0 & 0 & 0 & 8 & -32 & 56 & -96 & 224 & -384 \\
 1 & 0 & 0 & -2 & 20 & -58 & 100 & -186 & 332 \\
 2 & 0 & 0 & -4 & 8 & 28 & -96 & 132 & -216 \\
 3 & 0 & 0 & 2 & -20 & 38 & 20 & -102 & 136 \\
 4 & 0 & 0 & 0 & 8 & -56 & 104 & -24 & -96 \\
 5 & 0 & 0 & 0 & 0 & 20 & -120 & 218 & -104 \\
 6 & 0 & 0 & 0 & 0 & 0 & 40 & -220 & 392 \\
 7 & 0 & 0 & 0 & 0 & 0 & 0 & 70 & -364 \\
 8 & 0 & 0 & 0 & 0 & 0 & 0 & 0 & 112\\ \hline
\end{array}\\
&
\begin{array}{|c|rrrrrrrr|}\hline c(m,n,|r|) &
\multicolumn{8}{|c|}{m,n}\\ \hline
|r|  & 2,0 & 2,1 & 2,2 & 2,3 & 2,4 & 2,5 & 2,6 & 2,7 \\
\hline
 0 & 0 & 0 & 64 & -280 & 436 & -936 & 3186 & -5712 \\
 1 & 0 & 0 & -50 & 240 & -422 & 864 & -2696 & 4868 \\
 2 & 0 & 0 & 28 & -168 & 385 & -704 & 1710 & -3072 \\
 3 & 0 & 0 & -14 & 122 & -392 & 686 & -1040 & 1568 \\
 4 & 0 & 0 & 4 & -76 & 410 & -996 & 1304 & -1120 \\
 5 & 0 & 0 & 0 & 22 & -274 & 1194 & -2468 & 2608 \\
 6 & 0 & 0 & 0 & 0 & 75 & -776 & 3002 & -5648 \\
 7 & 0 & 0 & 0 & 0 & 0 & 200 & -1860 & 6672 \\
 8 & 0 & 0 & 0 & 0 & 0 & 0 & 455 & -3944 \\
 9 & 0 & 0 & 0 & 0 & 0 & 0 & 0 & 924\\ \hline
\end{array}\\
&\begin{array}{|c|rrrrrrrr|}\hline c(m,n,|r|) &
\multicolumn{8}{|c|}{m,n}\\ \hline
|r|  & 3,0 & 3,1 & 3,2 & 3,3 & 3,4 & 3,5 & 3,6 & 3,7 \\
\hline
 0 & 0 & 0 & 80 & -464 & 2480 & -8832 & 24848 & -66544 \\
 1 & 0 & 0 & -44 & 344 & -1996 & 7828 & -22954 & 60440 \\
 2 & 0 & 0 & -16 & -76 & 860 & -5048 & 17988 & -47496 \\
 3 & 0 & 0 & 38 & -172 & 354 & 1220 & -10670 & 34724 \\
 4 & 0 & 0 & -24 & 264 & -1256 & 2824 & 576 & -19848 \\
 5 & 0 & 0 & 6 & -172 & 1442 & -5876 & 11846 & -4988 \\
 6 & 0 & 0 & 0 & 44 & -844 & 5944 & -21188 & 39616 \\
 7 & 0 & 0 & 0 & 0 & 200 & -3172 & 19762 & -64060 \\
 8 & 0 & 0 & 0 & 0 & 0 & 696 & -9800 & 55904 \\
 9 & 0 & 0 & 0 & 0 & 0 & 0 & 2016 & -26116 \\
 10 & 0 & 0 & 0 & 0 & 0 & 0 & 0 & 5096\\ \hline
\end{array}\end{align*}}\begin{table}[h]
\caption{\small Coefficients $c(n,m,|r|)$ in the expansions
\eqref{q11exp} of $q_{11}$ and \eqref{q22exp} of $q_{22}$ up to
$p_1^7p_2^7$.}\label{t:schq1c}
\end{table}
\begin{table}{\fontsize{10pt}{12pt}\selectfont
\begin{align*}
&\begin{array}{|c|rrrrrrrr|}\hline c(m,n,|r|) &
\multicolumn{8}{|c|}{m,n}\\ \hline
|r|  & 4,0 & 4,1 & 4,2 & 4,3 & 4,4 & 4,5 & 4,6 & 4,7 \\
\hline
 0 & 0 & 0 & 160 & -2488 & 12544 & -62120 & 234830 & -710704 \\
 1 & 0 & 0 & -114 & 2036 & -11144 & 55876 & -216304 & 662636 \\
 2 & 0 & 0 & 52 & -1136 & 8101 & -41400 & 169788 & -539568 \\
 3 & 0 & 0 & -46 & 566 & -5598 & 28290 & -117820 & 388576 \\
 4 & 0 & 0 & 52 & -508 & 4786 & -23796 & 86960 & -265696 \\
 5 & 0 & 0 & -32 & 522 & -4916 & 27014 & -93240 & 234296 \\
 6 & 0 & 0 & 8 & -312 & 4135 & -29328 & 122769 & -326680 \\
 7 & 0 & 0 & 0 & 76 & -2086 & 22632 & -132506 & 466876 \\
 8 & 0 & 0 & 0 & 0 & 450 & -10224 & 95553 & -490488 \\
 9 & 0 & 0 & 0 & 0 & 0 & 1996 & -39810 & 333608 \\
 10 & 0 & 0 & 0 & 0 & 0 & 0 & 7195 & -130376 \\
 11 & 0 & 0 & 0 & 0 & 0 & 0 & 0 & 22168
\\ \hline
\end{array}\\
&\begin{array}{|c|rrrrrrrr|}\hline c(m,n,|r|) &
\multicolumn{8}{|c|}{m,n}\\ \hline
|r|  & 5,0& 5,1& 5,2 & 5,3 & 5,4 & 5,5 & 5,6 & 5,7 \\
\hline
 0 & 0 & 0 & 128 & -3104 & 33968 & -233760 & 1147472 & -4587808 \\
 1 & 0 & 0 & -82 & 2100 & -27150 & 197620 & -1027478 & 4194536 \\
 2 & 0 & 0 & 16 & -276 & 11668 & -112016 & 713004 & -3189184 \\
 3 & 0 & 0 & 12 & -680 & 2752 & 22236 & -320650 & 1932252 \\
 4 & 0 & 0 & -40 & 808 & -10712 & 44088 & -32320 & -706120 \\
 5 & 0 & 0 & 60 & -904 & 13504 & -86580 & 307774 & -406088 \\
 6 & 0 & 0 & -40 & 900 & -13388 & 109452 & -524960 & 1485948 \\
 7 & 0 & 0 & 10 & -516 & 9988 & -105040 & 650566 & -2512864 \\
 8 & 0 & 0 & 0 & 120 & -4552 & 70344 & -594360 & 3069848 \\
 9 & 0 & 0 & 0 & 0 & 906 & -28236 & 367760 & -2675536 \\
 10 & 0 & 0 & 0 & 0 & 0 & 5012 & -135100 & 1552196 \\
 11 & 0 & 0 & 0 & 0 & 0 & 0 & 22028 & -532300 \\
 12 & 0 & 0 & 0 & 0 & 0 & 0 & 0 & 81216
\\ \hline
\end{array}\\
&\begin{array}{|c|rrrrrrrr|}\hline c(m,n,|r|) &
\multicolumn{8}{|c|}{m,n}\\ \hline
|r| & 6,0& 6,1 & 6,2 & 6,3 & 6,4 & 6,5 & 6,6 & 6,7 \\
\hline
 0 & 0 & 0 & 280 & -10872 & 102536 & -1138984 & 5877076 &
   -31532632 \\
 1 & 0 & 0 & -188 & 8728 & -87570 & 996452 & -5334766 & 28957580
   \\
 2 & 0 & 0 & 40 & -4692 & 55878 & -676608 & 3989676 & -22456872 \\
 3 & 0 & 0 & 22 & 2144 & -31748 & 387288 & -2527130 & 14835288 \\
 4 & 0 & 0 & -8 & -1352 & 25320 & -249096 & 1597629 & -8860904 \\
 5 & 0 & 0 & -42 & 1304 & -29168 & 249912 & -1429278 & 6205936 \\
 6 & 0 & 0 & 72 & -1548 & 33081 & -313176 & 1838208 & -7147328 \\
 7 & 0 & 0 & -48 & 1470 & -31172 & 353622 & -2414106 & 10644034 \\
 8 & 0 & 0 & 12 & -796 & 21444 & -309284 & 2630054 & -14314652 \\
 9 & 0 & 0 & 0 & 178 & -9014 & 188102 & -2141792 & 15064826 \\
 10 & 0 & 0 & 0 & 0 & 1681 & -69096 & 1191380 & -11542392 \\
 11 & 0 & 0 & 0 & 0 & 0 & 11376 & -398304 & 5977772 \\
 12 & 0 & 0 & 0 & 0 & 0 & 0 & 59891 & -1857136 \\
 13 & 0 & 0 & 0 & 0 & 0 & 0 & 0 & 260164\\ \hline
\end{array}
\end{align*}}
\caption{\small Coefficients $c(n,m,|r|)$ in the expansions
\eqref{q11exp} of $q_{11}$ and \eqref{q22exp} of $q_{22}$ up to
$p_1^7p_2^7$.}\label{t:schq1d}
\end{table}

\begin{table}{\fontsize{10pt}{12pt}\selectfont
\begin{align*}
&\begin{array}{|c|rrrrrrrrrrrr|}\hline d(m,n,r) &
\multicolumn{12}{|c|}{m,n}
\\ \hline
r& 1,1& 1,2& 1,3& 2,2& 1,4& 2,3&
1,5& 2,4& 3,3& 1,6 &2,5 & 3,4\\
\hline
-7 & 0 & 0 & 0 & 0 & 0 & 0 & 0 & 0 & 0 & -2 & -18 & -60 \\
 -6 & 0 & 0 & 0 & 0 & 0 & 0 & -2 & -11 & -22 & 4 & 48 & 220 \\
 -5 & 0 & 0 & 0 & 0 & -2 & -6 & 4 & 24 & 72 & -2 & -48 & -374 \\
 -4 & 0 & 0 & -2 & -3 & 4 & 8 & -2 & -21 & -158 & -2 & 24 & 300 \\
 -3 & 0 & -2 & 4 & -4 & -4 & -10 & 0 & -8 & 268 & 2 & -48 & 130 \\
 -2 & -2 & 2 & -4 & 24 & 2 & 48 & -2 & 89 & -174 & 0 & 188 & -336
   \\
 -1 & 4 & 2 & 4 & -16 & 2 & -60 & 4 & -100 & -132 & 2 & -242 &
   -188 \\
 0 & 0 & 0 & 0 & -20 & 0 & -8 & 0 & -60 & 232 & 0 & 8 & 608 \\
 1 & -4 & -2 & -4 & 32 & -2 & 44 & -4 & 164 & -76 & -2 & 178 &
   -312 \\
 2 & 2 & -2 & 4 & -8 & -2 & -16 & 2 & -85 & -46 & 0 & -100 & -52
   \\
 3 & 0 & 2 & -4 & -12 & 4 & 18 & 0 & -16 & 12 & -2 & 24 & 130 \\
 4 & 0 & 0 & 2 & 7 & -4 & -32 & 2 & 65 & 118 & 2 & -64 & -300 \\
 5 & 0 & 0 & 0 & 0 & 2 & 14 & -4 & -64 & -144 & 2 & 120 & 546 \\
 6 & 0 & 0 & 0 & 0 & 0 & 0 & 2 & 23 & 50 & -4 & -104 & -440 \\
 7 & 0 & 0 & 0 & 0 & 0 & 0 & 0 & 0 & 0 & 2 & 34 & 128
\\ \hline
\end{array}\\
&
\begin{array}{|c|rrrrrrrrrr|}\hline d(m,n,r) &
\multicolumn{10}{|c|}{m,n}
\\ \hline
r & 1,7& 2,6 & 3,5 & 4,4 & 2,7 & 3,6 & 4,5 & 3,7 & 4,6 & 5,5 \\
\hline
 -10 & 0 & 0 & 0 & 0 & 0 & 0 & 0 & -466 & -1526 & -2232 \\
 -9 & 0 & 0 & 0 & 0 & -38 & -262 & -624 & 2032 & 7268 & 10940 \\
 -8 & -2 & -27 & -134 & -218 & 120 & 1092 & 2766 & -3758 & -14515
   & -22682 \\
 -7 & 4 & 80 & 528 & 888 & -144 & -1940 & -5026 & 4020 & 15684 &
   26800 \\
 -6 & -2 & -90 & -902 & -1448 & 84 & 2010 & 4624 & -3098 & -9540 &
   -24552 \\
 -5 & 0 & 48 & 912 & 800 & -38 & -1558 & -2624 & 2092 & -200 &
   28056 \\
 -4 & 0 & -38 & -918 & 896 & 32 & 610 & 2852 & -2272 & 13280 &
   -35622 \\
 -3 & 0 & 24 & 1612 & -1340 & -110 & 1950 & -3284 & 6156 & -20848
   & 37916 \\
 -2 & -2 & 158 & -1990 & -368 & 412 & -4210 & -904 & -10692 & 5868
   & -37754 \\
 -1 & 4 & -232 & 372 & 1624 & -538 & 1734 & 5050 & 6364 & 21792 &
   29244 \\
 0 & 0 & -124 & 1664 & -1357 & 24 & 3412 & -2460 & 4704 & -30526 &
   2220 \\
 1 & -4 & 368 & -1572 & 1168 & 426 & -4798 & -3038 & -8788 & 16792
   & -38148 \\
 2 & 2 & -158 & 250 & -1176 & -276 & 2410 & 5588 & 4140 & -4608 &
   43578 \\
 3 & 0 & -32 & 284 & 716 & 62 & -438 & -5176 & -276 & 4952 &
   -19812 \\
 4 & 0 & 46 & 206 & 320 & -32 & -274 & 2700 & 224 & -8096 & -3074
   \\
 5 & 0 & -104 & -1096 & -1968 & 54 & 1326 & 2684 & -1444 & 2176 &
   2208 \\
 6 & 2 & 186 & 1554 & 2880 & -156 & -2962 & -9148 & 3630 & 15114 &
   24366 \\
 7 & -4 & -152 & -1040 & -1888 & 264 & 3484 & 10694 & -6404 &
   -31876 & -52300 \\
 8 & 2 & 47 & 270 & 471 & -208 & -2088 & -6018 & 6734 & 31313 &
   50774 \\
 9 & 0 & 0 & 0 & 0 & 62 & 502 & 1344 & -3752 & -15740 & -24904 \\
 10 & 0 & 0 & 0 & 0 & 0 & 0 & 0 & 854 & 3236 & 4978\\ \hline
\end{array}
\end{align*}}\caption{\small Coefficients $d(n,m,r)$ in the expansions
\eqref{q12exp} of $q_{12}$ up to $p_1^7p_2^7$.}\label{t:schzb}
\end{table}

\begin{table}{\fontsize{10pt}{12pt}\selectfont
$$
\begin{array}{|c|rrrrrr|}\hline d(m,n,r) &
\multicolumn{6}{|c|}{m,n}
\\ \hline
r & 4,7 & 5,6& 5,7& 6,6& 6,7& 7,7\\
\hline
 -14 & 0 & 0 & 0 & 0 & 0 & -279508 \\
 -13 & 0 & 0 & 0 & 0 & -75924 & 1927676 \\
 -12 & 0 & 0 & -17630 & -24228 & 481740 & -5998022 \\
 -11 & -3330 & -6694 & 100956 & 142216 & -1364286 & 11228868 \\
 -10 & 16836 & 35728 & -253434 & -368166 & 2304584 & -14286298 \\
 -9 & -36242 & -82456 & 373560 & 563140 & -2636708 & 13345304 \\
 -8 & 43556 & 110530 & -375412 & -581504 & 2209596 & -10053096 \\
 -7 & -31916 & -102420 & 303168 & 424564 & -1441188 & 7731048 \\
 -6 & 13316 & 77064 & -265734 & -156462 & 840784 & -8536940 \\
 -5 & -5756 & -46470 & 334836 & -162908 & -670488 & 11904316 \\
 -4 & 25424 & 16218 & -467020 & 433181 & 841880 & -14909588 \\
 -3 & -53870 & 3862 & 516756 & -506252 & -1008406 & 14283736 \\
 -2 & 34092 & -27582 & -410996 & 240120 & 785932 & -8781634 \\
 -1 & 32538 & 48822 & 184292 & 293204 & -164092 & 221276 \\
 0 & -62216 & -14228 & 127836 & -707758 & -289064 & 7537956 \\
 1 & 21398 & -63242 & -399996 & 725940 & 122532 & -10976964 \\
 2 & 30028 & 94466 & 436396 & -500264 & 368740 & 9267180 \\
 3 & -45058 & -65554 & -247420 & 288796 & -681688 & -4644648 \\
 4 & 35136 & 41606 & 31570 & -175530 & 734824 & 164070 \\
 5 & -18408 & -31580 & 81704 & 137264 & -692102 & 2588888 \\
 6 & -10816 & -25534 & -34490 & -23790 & 468580 & -3291630 \\
 7 & 56288 & 143206 & -232388 & -378460 & 511608 & 850620 \\
 8 & -89580 & -225998 & 614472 & 953710 & -2611364 & 6910004 \\
 9 & 77454 & 187450 & -784572 & -1191972 & 4740264 & -18914964 \\
 10 & -35776 & -82270 & 574978 & 854578 & -4995180 & 27421428 \\
 11 & 6902 & 15076 & -230896 & -335532 & 3183372 & -24917120 \\
 12 & 0 & 0 & 39464 & 56113 & -1141052 & 14172292 \\
 13 & 0 & 0 & 0 & 0 & 177106 & -4628036 \\
 14 & 0 & 0 & 0 & 0 & 0 & 663786\\ \hline
\end{array}
$$}\caption{\small Coefficients $d(n,m,r)$ in the expansions
\eqref{q12exp} of $q_{12}$ up to $p_1^7p_2^7$.}\label{t:schzc}
\end{table}

\begin{table}{\fontsize{10pt}{12pt}\selectfont
\begin{align*}&\begin{array}{|c|rrrrrrrr|}\hline b(m,n,|r|) &
\multicolumn{8}{|c|}{m,n}
\\ \hline
|r|& 0,0& 0,1& 0,2& 0,3&  0,4& 0,5& 0,6 & 0,7\\ \hline 0& 1 & -24 &
252 & -1472 & 4830 & -6048 & -16744 & 84480 \\ \hline
\end{array}\\
&
\begin{array}{|c|rrrrrrrrr|}
 \hline b(m,n,|r|) & \multicolumn{9}{|c|}{m,n}\\
 \hline
 |r| & 1,1& 1,2& 1,3& 2,2& 1,4& 2,3& 1,5& 2,4& 3,3\\
\hline 0 & 528 & -4944 & 24336 & 41640 & -56256 & -205632 & -27024 &
   816288 & 1713888 \\
 1 & 48 & -1152 & 12144 & 24672 & -71808 & -216816 & 243984 &
   883104 & 1296576 \\
 2 & -24 & 648 & -7872 & -15480 & 55848 & 160032 & -247848 &
   -889488 & -1286808 \\
 3 & 0 & -48 & 1296 & 1680 & -15696 & -24480 & 110400 & 189696 &
   143728 \\
 4 & 0 & 0 & -72 & 60 & 1920 & -1920 & -22872 & 32124 & 102480 \\
 5 & 0 & 0 & 0 & 0 & -96 & 528 & 2544 & -16560 & -32304 \\
 6 & 0 & 0 & 0 & 0 & 0 & 0 & -120 & 1560 & 2776
\\ \hline
\end{array}\\
&\begin{array}{|c|rrrrrrr|}
 \hline b(m,n,|r|) & \multicolumn{7}{|c|}{m,n}\\
 \hline |r|& 1,6 &2,5 & 3,4 & 1,7 & 2,6 & 3,5 & 4,4\\
\hline
 0 & 519984 & -3852288 & -14764128 & -1105680 & 18817224 &
   92719872 & 145219728 \\
 1 & -362112 & -318384 & 366192 & -559728 & -14508240 & -46776288
   & -63785808 \\
 2 & 652944 & 2391648 & 4180872 & -609336 & 1574328 & 6267336 &
   7050864 \\
 3 & -480000 & -747120 & 395760 & 1195488 & 144720 & -10784880 &
   -19196592 \\
 4 & 157176 & -360384 & -1724592 & -658896 & 2877768 & 16216944 &
   24936900 \\
 5 & -30096 & 242880 & 724128 & 205248 & -2189376 & -8860944 &
   -12821664 \\
 6 & 3168 & -47904 & -122472 & -37320 & 677976 & 2303352 & 3237528
   \\
 7 & -144 & 3360 & 7296 & 3792 & -101664 & -287904 & -380784 \\
 8 & 0 & 0 & 0 & -168 & 6132 & 13776 & 14142\\ \hline
\end{array}\\
&\begin{array}{|c|rrrrrr|}
 \hline b(m,n,|r|) & \multicolumn{6}{|c|}{m,n}\\
 \hline |r|& 2,7 &3,6 & 4,5 & 3,7 & 4,6 & 5,5 \\
 \hline 0 & -66554496 & -358727760 & -726665184 & 695673744 & 1351136784 & 1650790560 \\
 1 & 75180960 & 274075344 & 428621472 & -674344080 & -718078080 &
   -483068160 \\
 2 & -38527584 & -102971760 & -102840216 & 311462112 & -280279080
   & -862545528 \\
 3 & 14950992 & 68820304 & 117592080 & -183732528 & 76947936 &
   326955216 \\
 4 & -16461504 & -95548032 & -184485168 & 341473632 & 669734904 &
   803199888 \\
 5 & 13341840 & 67657776 & 126979872 & -332188944 & -746531664 &
   -953493840 \\
 6 & -5799072 & -24827408 & -44557992 & 169272720 & 363331344 &
   460103064 \\
 7 & 1409568 & 4971792 & 8017536 & -49583184 & -91365264 &
   -107659296 \\
 8 & -183552 & -510936 & -603840 & 8326104 & 10271160 & 8600928 \\
 9 & 10080 & 20384 & 2928 & -723168 & 42144 & 969888 \\
 10 & 0 & 0 & 0 & 23184 & -78552 & -168288\\ \hline
\end{array}\\
&\begin{array}{|c|rrrr|}
 \hline b(m,n,|r|) & \multicolumn{4}{|c|}{m,n}\\
 \hline |r|& 4,7 &5,6 &  5,7 & 6,6  \\
\hline
 0 & 4832096064 & 10189344912 & -84307019856 & -101975119376 \\
 1 & -5440894848 & -13330472448 & 95341959696 & 121249770048 \\
 2 & 7092433776 & 16964315712 & -112768635696 & -148479046944 \\
 3 & -5324456688 & -12179317152 & 99096258672 & 130813223840 \\
 4 & 493050816 & 2220117600 & -50945277360 & -67115756268 \\
 5 & 2182452480 & 3136424112 & 9275595696 & 12522143184 \\
 6 & -1723992024 & -2526256824 & 3218889000 & 3095832920 \\
 7 & 596460192 & 708556608 & -511386144 & 697588224 \\
 8 & -83802720 & 1655064 & -1637032632 & -2973944136 \\
 9 & -6562944 & -49812096 & 1112460432 & 1741704736 \\
 10 & 3622248 & 11648184 & -331421688 & -490584408 \\
 11 & -339120 & -897360 & 49702752 & 71074800 \\
 12 & 0 & 0 & -3070320 & -4265540\\ \hline
\end{array}
\end{align*}}\caption{\small Coefficients $b(n,m,|r|)$ of the function \eqref{zogF} up to $p_1^7p_2^7$.}\label{t:schfua}
\end{table}

\begin{table}{\fontsize{10pt}{12pt}\selectfont
\begin{align*}
& \begin{array}{|c|rr|}
 \hline b(m,n,|r|) & \multicolumn{2}{|c|}{m,n}\\
 \hline |r|& 6,7 & 7,7  \\
\hline
 0 & 172479560352 & 3128606566176 \\
 1 & -310507917168 & -2134241509440 \\
 2 & 580363077120 & -86143852200 \\
 3 & -687760594368 & 1889752759536 \\
 4 & 507099802584 & -2172090239616 \\
 5 & -235110351024 & 1435905509328 \\
 6 & 90844794528 & -861256923216 \\
 7 & -63838287456 & 741491756592 \\
 8 & 53462494704 & -656667020112 \\
 9 & -30023682864 & 420804461904 \\
 10 & 10521776736 & -181869137376 \\
 11 & -2257032864 & 52175288688 \\
 12 & 273215976 & -9563257536 \\
 13 & -14342640 & 1014591120 \\
 14 & 0 & -47275560\\ \hline
\end{array}
\end{align*}}\caption{\small Coefficients $b(n,m,|r|)$ of the function \eqref{zogF} up to $p_1^7p_2^7$.}\label{t:schfub}
\end{table}

\end{document}